\documentclass[preprint,showpacs,nofootinbib,amsmath,floatfix]{revtex4}

\usepackage{times}
\usepackage{hyperref}

\usepackage{graphicx}
\usepackage{dcolumn}
\usepackage{bm}

\begin{document}

\title{Radiative corrections to the Dalitz plot of $K_{l3}^\pm$ decays
}

\author{
C.\ Ju\'arez-Le\'on
}
\affiliation{
Departamento de F{\'\i}sica, Escuela Superior de F\'{\i}sica y Matem\'aticas del IPN, Apartado Postal 75-702, M\'exico, D.F.\ 07738, Mexico
}

\author{
A.\ Mart{\'\i}nez
}
\affiliation{
Departamento de F{\'\i}sica, Escuela Superior de F\'{\i}sica y Matem\'aticas del IPN, Apartado Postal 75-702, M\'exico, D.F.\ 07738, Mexico
}

\author{
M.\ Neri
}
\affiliation{
Departamento de F{\'\i}sica, Escuela Superior de F\'{\i}sica y Matem\'aticas del IPN, Apartado Postal 75-702, M\'exico, D.F.\ 07738, Mexico
}

\author{
J.\ J.\ Torres
}
\affiliation{
Departamento de Posgrado, Escuela Superior de C\'omputo del IPN, Apartado Postal 75-702, M\'exico, D.F.\ 07738, Mexico
}

\author{
Rub\'en Flores-Mendieta
}
\affiliation{
Instituto de F{\'\i}sica, Universidad Aut\'onoma de San Luis Potos{\'\i}, \'Alvaro Obreg\'on 64, Zona Centro, San Luis Potos{\'\i}, S.L.P.\ 78000, Mexico
}

\date{\today}

\begin{abstract}
We calculate the model-independent radiative corrections to the Dalitz plot of $K_{l3}^\pm$ decays to the order of $(\alpha/\pi)(q/M_1)$, where $q$ is the momentum transfer and $M_1$ is the mass of the kaon. The final results are presented, first, with the triple integration over the variables of the bremsstrahlung photon ready to be performed numerically and, second, in an analytical form. These two forms are useful to cross-check on one another and with other calculations. This paper is organized to make it accessible and reliable in the analysis of the Dalitz plot of precision experiments and is not compromised to fixing the form factors at predetermined values. It is assumed that the real photons are kinematically discriminated. Otherwise, our results have a general model-independent applicability.
\end{abstract}


\maketitle

\section{Introduction}

Analyses of the low-energy $s\to u$ and $d\to u$ semileptonic transitions play a decisive role in our understanding of the interplay between weak and strong interactions and the Cabibbo-Kobayashi-Maskawa (CKM) quark-mixing matrix. Current determinations of $|V_{ud}|$ and $|V_{us}|$ provide the most precise constraints on the size of the CKM matrix elements and yield the most stringent test of CKM unitarity. In particular, the best determination of $|V_{us}|$ is achieved from kaon semileptonic ($K_{l3}$) and leptonic ($K_{l2}$) decays, whereas the most precise determination of $|V_{ud}|$ is obtained from superallowed $0^+\to 0^+$ Fermi transitions and also, to a minor extent, from baryon semileptonic decays --most commonly from neutron beta decay-- and pion beta decay.

Over the past years, precise measurements have been made in both kaon and baryon semileptonic decays \cite{part}. In these experiments the statistical errors are rather small, and more effort has been put into the reduction of the systematic errors, which are mainly of two kinds. The first one comes from the different shortcomings of the experimental devices. The second one, of a theoretical nature, comprises assumptions about form factors and radiative corrections.

On the one hand, while the leptonic part of $s\to u$ and $d\to u$ semileptonic transitions is unambiguous, the hadronic part is affected by flavor SU(3) symmetry breaking in the form factors. For $K_{l3}$ decays there is a remarkable simplification because only the vector part of the weak current has a nonvanishing contribution and only two form factors appear. These form factors are protected by the Ademollo-Gatto theorem \cite{ag} against SU(3) breaking corrections to lowest order in $(m_s-\hat{m})$ so that the theoretical approach to compute them is under reasonable control within the limits of experimental precision. In contrast, baryon semileptonic decays play an ancillary role due to the presence of both vector and axial-vector currents and the appearance of six form factors. Although the leading vector form factor is also protected by the Ademollo-Gatto theorem, large theoretical uncertainties arise from first-order SU(3) breaking effects in the axial-vector form factors.

On the other hand, the precise and reliable calculation of radiative corrections to the various measurable quantities accessible to current experiments is a rather difficult task. Despite the enormous progress achieved in the understanding of the fundamental interactions with the standard model, no first-principles calculation of radiative corrections to kaon and baryon semileptonic decays is yet possible. Thus, in dealing with these radiative corrections it is very important to have an organized systematic approach to deal with the complications and technical difficulties that accompany them. These radiative corrections depend on the details of strong and weak interactions present in the decay vertex; i.e., they have a model-dependent part. They also depend on the charge assignments of the decaying and emitted hadrons. Furthermore, their final form depends on the observed kinematical and angular variables and on certain experimental conditions.

Various articles on radiative corrections to $K_{l3}$ decays exist in the literature, each with a different emphasis. Some of the earliest attempts can be traced back to the works by Ginsberg \cite{ginsKl3pm,ginsKe3pmDP,ginsKe30,ginsKm3}, Becherrawy \cite{bech}, Garc{\'\i}a and Maya \cite{maya}, and more recently by Cirigliano \textit{et.al.} \cite{cir,cir2}, Bytev \textit{et.\ al.} \cite{bytev}, and Andre \cite{andre}, to name but a few. Ginsberg, for instance, focused on the calculation of radiative corrections to the lepton spectrum, Dalitz plot, and decay rates of $K_{l3}$ decays, by assuming a phenomenological weak $K$-$\pi$ vertex; the results depend on a cutoff. Becherrawy, in turn, used a particular model of the strong interactions, whereas Garc{\'\i}a and Maya extended to $M_{l3}$ the procedure proposed by Sirlin \cite{sirlin}, originally introduced to study the radiative corrections to the charged lepton spectrum in neutron beta decay. Cirigliano \textit{et.\ al.} computed these radiative corrections in the framework of the chiral perturbation theory and accounted for virtual photons and leptons. Bytev \textit{et.\ al.}, in turn, presented an alternative approach to remove the ultraviolet cutoff dependence and set it equal to the $W$ mass. Finally, Andre focused his analysis on $K_{l3}^0$ decays and included contributions from outside the kinematically allowed three-body region of the Dalitz plot.

As for baryon semileptonic decays, following the analysis of Sirlin of the virtual radiative corrections in neutron beta decay \cite{sirlin} and armed with the theorem of Low for the bremsstrahlung radiative corrections \cite{low,chew}, it has been shown that the model dependence that arises in this latter case contributes to the order of $(\alpha/\pi)(q/M_B)^n$ for $n\geq 2$ and that such model dependence can be absorbed into the already existing form factors \cite{juarez}, while the model-independent contributions can be extracted from orders $n=0,1$. Here $M_B$ is the mass of the decaying baryon, and $q$ is the four-momentum transfer. This approach has been implemented to compute high-precision radiative corrections to various observables in baryon semileptonic decays (decay rates and angular spin-asymmetry coefficients) by considering unpolarized and polarized decaying and emitted baryons \cite{jl2009,jj2006}, for all possible charge assignments.

In this paper, we reexamine the calculation of radiative corrections to $K_{l3}^\pm$ decays up to the order of $(\alpha/\pi)(q/M_1)$, where $M_1$ denotes the mass of the kaon. Our analysis builds on earlier works, particularly Ref.~\cite{maya}, but unlike it, here we will focus on the radiative corrections to the Dalitz plot of $K_{l3}^\pm$ decays. Our approach to the calculation of these radiative corrections has been to advance results which can be established as much as possible once and for all. This task is considerably biased by the experimental precision attained in given experiments and by the available phase space in each decay. The radiative corrections obtained to the order of approximation considered here are then suitable for model-independent experimental analyses and are valid to an acceptable degree of precision. We assume that the form factors can be extracted from experiment, and thus we do not consider them here. Their model-independent parts contain information only on QED.

Our results will be presented in two forms: one where the triple integration over the real photon three-momentum is left indicated and ready to be performed numerically and another one, of analytical nature, where such an integration has been performed. Both forms can be used to numerically cross-check one another. However, the analytical result, although tedious to feed into a Monte Carlo program, can help in the reduction of computer time because the triple integration involved does not have to be performed within the Monte Carlo calculation every time the energies of the charged lepton and the pion or the form factors are changed.

This paper in organized as follows. In Sec.~\ref{sec:thkl3}, we introduce our notation and conventions; also we briefly discuss some basic aspects of the phenomenology of kaon semileptonic decays and define the boundaries of the allowed kinematical region, usually referred to as the Dalitz plot. We will specialize our calculation to the three-body region of this plot. We proceed to analyze in Sec.~\ref{sec:vir} the virtual radiative corrections and properly identify the infrared-divergent term. In Sec.~\ref{sec:bremss}, we discuss the bremsstrahlung radiative corrections and also extract the infrared divergence and the finite terms that accompany it. We are left with several triple integrals, mainly over the photon variables, which at any rate can be computed numerically. We, however, proceed further in order to analytically perform these integrals, and we present the resultant expressions. We also present some cross-checks for completeness. We summarize our findings in Sec.~\ref{sec:close}, where we also carry out a detailed numerical evaluation of these radiative corrections at several points of the Dalitz plot; this way we are able to compare our results with others already published, in particular, with Refs.~\cite{cir,cir2}. We close this paper with a brief discussion of our results and some concluding remarks in Sec.~\ref{sec:finalsec}. This work is complemented with two appendixes. In Appendix \ref{app:drbr}, we discuss some features on how to deal with the bremsstrahlung differential decay rate in order to properly isolate the infrared divergence. In Appendix \ref{app:tripleint}, we list the triple integrals over the photon variables that emerge in our calculation.

\section{\label{sec:thkl3}Basics on kaon semileptonic decays}

In this section, we provide a basic description of the phenomenology about kaon semileptonic decays in order to introduce our notation and conventions.

For definiteness, we concentrate on the analysis of the semileptonic decay of a positively charged kaon, namely,
\begin{equation}
K^+(p_1) \to \pi^0(p_2) + \ell^+(l) + \nu_\ell (p_\nu^0), \label{eq:kl3p}
\end{equation}
because the corresponding charge conjugate mode can be analyzed in a similar way. The four-momenta and masses of the $K^+$, $\pi^0$, $\ell^+$, and $\nu_\ell$ will be denoted by $p_1=(E_1,{\mathbf p}_1)$, $p_2=(E_2,{\mathbf p}_2)$, $l=(E,{\mathbf l})$, and $p_\nu=(E_\nu^0,{\mathbf p}_\nu^0)$, and by $M_1$, $M_2$, $m$, and $m_\nu$, respectively. Because no assumptions will be made about the size of $m$ compared to $M_1$, our results will be valid and applicable to both $K_{e3}^\pm$ and $K_{\mu 3}^\pm$ decays. Without loss of generality, the reference system we will use throughout this paper is the center-of-mass frame, which corresponds to the rest frame of $K^+$, so that all expressions which are not manifestly covariant will apply to this frame. In this regard, quantities like $p_2$, $l$, or $p_\nu$ will also denote the magnitudes of the corresponding three-momenta, unless explicitly stated otherwise. In passing, let us mention that the direction of a generic three-vector ${\mathbf p}$ will be denoted by a unit vector $\hat {\mathbf p}$.

\begin{figure}
{\includegraphics{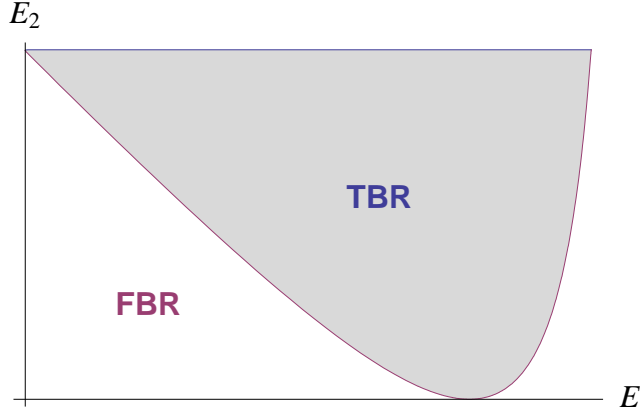}
\begin{center}
(a)
\end{center}
\includegraphics{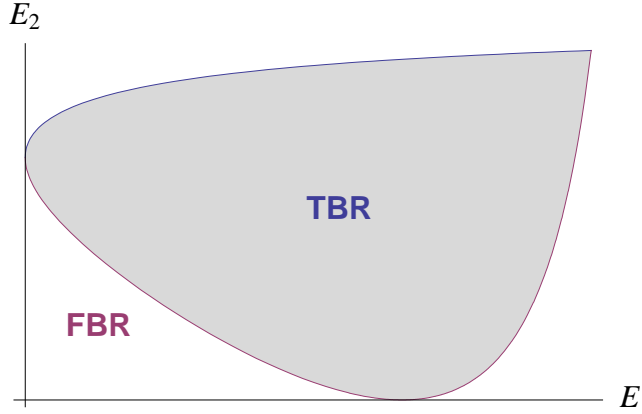}
\begin{center}
(b)
\end{center}}
\caption{\label{fig:kinem} Kinematical region for $K_{l3}^\pm$ decays as a function of the energies of the charged lepton and the emitted pion, $E$ and $E_2$, respectively, for (a) $K_{e3}^\pm$ and (b) $K_{\mu 3}^\pm$ decays. In this work, areas TBR and FBR are loosely referred to as the three- and four-body regions of the Dalitz plot, respectively.}
\end{figure}

The allowed kinematical region in the variables $E$ and $E_2$ for kaon semileptonic decays, usually referred to as the Dalitz plot, is determined by energy and momentum conservation. This region is depicted in Fig.~\ref{fig:kinem}, in which we have distinguished two areas labeled as TBR (three-body region) and FBR (four-body region).
The former (the shaded region in this figure) is bounded by
\begin{equation}
E_2^{\mathrm{min}} \leq E_2 \leq E_2^{\textrm{max}}, \qquad \quad m \leq E \leq E_m, \label{eq:EE2lim}
\end{equation}
where
\begin{equation}
E_2^{\mathrm{max,min}} = \frac12 (M_1-E \pm l) + \frac{M_2^2}{2(M_1-E\pm l)},
\end{equation}
and
\begin{equation}
E_m = \frac{1}{2M_1} (M_1^2-M_2^2+m^2), \label{eq:Elim}
\end{equation}
whereas the latter is delimited by
\begin{equation}
M_2 \leq E_2 \leq E_2^{\textrm{min}}, \qquad \quad m \leq E \leq E_c,
\end{equation}
where
\begin{equation}
E_c = \frac12 (M_1-M_2) + \frac{m^2}{2(M_1-M_2)}.
\end{equation}

The distinction between these two areas is subtle and requires some discussion. First, let us notice that in order to find an event with energies $E$ and $E_2$ in area FBR a fourth particle must exist (in our case, such a particle will be a photon) which will carry away finite energy and momentum. In contrast, in area TBR this particle may or may not do so. It turns out that area FBR is exclusively a four-body region, whereas area TBR is both a three- and a four-body region. Hereafter, areas TBR and FBR will be loosely referred to as the three- and four-body regions of the Dalitz plot, respectively.

We can now proceed to the construction of the uncorrected transition amplitude (i.e., the amplitude without radiative corrections) for process~(\ref{eq:kl3p}). By neglecting scalar and tensor contributions and assuming that only the vector current contributes, such an amplitude, denoted here by $\mathsf{M}_0$, has the structure
\begin{eqnarray}
\mathsf{M}_0 = C_K \frac{G_F}{\sqrt 2} V_{us}^* W_\mu(p_1,p_2) \left[\overline{u}_\nu(p_\nu) O_\mu v_\ell(l)\right],
\label{eq:M0}
\end{eqnarray}
where $G_F$ is the Fermi constant as extracted from muon decay, $V_{us}$ is the relevant CKM matrix element, and $C_K=1/\sqrt{2}$ is a Clebsch-Gordan coefficient. In order to avoid cumbersome expressions, the factors $G_F$, $V_{us}$ and $C_K$ will be included into a single factor $G$ hereafter. Furthermore, $v_\ell$ and $u_\nu$ are the Dirac spinors of the corresponding particles, $O_\mu \equiv \gamma_\mu(1+\gamma_5)$, and the metric and $\gamma$-matrix convention we adopt is the standard one (see, for instance, Ref.~\cite{bd}), except that our $\gamma_5$ has the opposite sign.

The hadronic matrix element $W_\mu(p_1,p_2)$, on the other hand, can be written as
\begin{eqnarray}
W_\mu (p_1,p_2) & = & \langle \pi^0(p_2)|\bar{u} \gamma_\mu s|K^+(p_1) \rangle \nonumber \\
& = & f_+(q^2) (p_1+p_2)_\mu + f_-(q^2) (p_1-p_2)_\mu. \label{eq:mtxel}
\end{eqnarray}
Here $q=p_1-p_2$ is the four-momentum transfer and $f_\pm (q^2)$ are dimensionless form factors which are relatively real if time reversal holds.

Let us remark that the momentum transfer dependence of the form factors has been discussed extensively in the literature; in particular, Ref.~\cite{part} provides an overview about the current situation of both the theoretical and experimental bent. In summary, a common practice advocated in $K_{\mu 3}$ analyses is the use of a linear parametrization such as
\begin{equation}
f_\pm(q^2) = f_\pm(0) \left[1+\lambda_\pm \frac{q^2}{M_2^2} \right], \label{eq:ffq2}
\end{equation}
where $\lambda_\pm$ are slope parameters. Most $K_{\mu 3}$ data are well described by this parametrization for $\lambda_-=0$ \cite{part}. Recent analyses, however, have opted to use the $(\lambda_+,\lambda_0)$ parametrization instead, which is based on the introduction of an alternative set of form factors, namely,
\begin{equation}
f_0(q^2) = f_+(q^2) + \frac{q^2}{M_1^2-M_2^2} f_-(q^2),
\end{equation}
where the vector and scalar form factors $f_+(q^2)$ and $f_0(q^2)$ represent the $p$ and $s$ wave projections of the crossed channel $\langle 0|\bar{u} \gamma_\mu s|K\pi \rangle$, respectively. Again, if it is assumed that $f_+$ follows a linear parametrization and $f_-$ is constant, then
\begin{equation}
f_0(q^2) = f_0(0) \left[1 + \lambda_0 \frac{q^2}{M_2^2} \right].
\end{equation}

Finally, another parametrization which has gained a renewal of interest is the dispersive parametrization \cite{bernard}, which is based on dispersive techniques and the low-energy $K$-$\pi$ phases to parametrize the form factors, both scalar and vector.

As we have pointed out in the introductory section, we assume that the form factors are determined from experiment and will not be determined here. Therefore, once the decay amplitude $\mathsf{M}_0$ is defined the uncorrected differential decay rate for process (\ref{eq:kl3p}), represented here by $d\Gamma_0$, can readily be obtained by standard techniques. The differential decay rate is thus given by \cite{bd}
\begin{equation}
d\Gamma_0 = \frac{1}{2M_1} \frac{d^3p_2}{2E_2(2\pi)^3} \frac{m}{E} \frac{d^3l}{(2\pi)^3} \frac{m_\nu}{E_\nu^0} \frac{d^3p_\nu^0}{(2\pi)^3} (2\pi)^4 \delta^4(p_1-p_2-l-p_\nu^0)\sum_{\textrm{spins}}|\mathsf{M}_0|^2.
\end{equation}
The integral over the three-momentum of the neutrino is straightforward. Besides, the integrals over the angular variables facilitate if, without loss of generality, we orient the coordinate axes in such a way that $\ell^+$ is emitted along the $+z$ axis and $\pi^0$ is emitted in the first or fourth quadrant of the $(x,z)$ plane. Then, the only nontrivial angular integration is over $\theta_2$, namely
\begin{equation}
d\Gamma_0 = \frac{1}{(2\pi)^3}\frac{mm_\nu}{2M_1}dEdE_2\int_{-1}^1dy\delta(y-y_0) \sum_{\textrm{spins}}|\mathsf{M}_0|^2,
\end{equation}
where $y=\cos\theta_2$ and
\begin{equation}
y_0 = \frac{{E_\nu^0}^2-p_2^2-l^2}{2p_2l}. \label{eq:y0}
\end{equation}

After some rearrangements, the resultant expression can be cast into the compact form
\begin{equation}
d\Gamma_0 = A_0 \, d\Omega, \label{eq:drate0}
\end{equation}
where $A_0$ is a function of the kinematical variables and depends quadratically on the form factors. It can be organized as
\begin{equation}
A_0 = A_1^{(0)} |f_+(q^2)|^2 + A_2^{(0)} \mathrm{Re} \,[f_+(q^2) f_-^*(q^2)] + A_3^{(0)} |f_-(q^2)|^2, \label{eq:a0}
\end{equation}
with
\begin{equation}
A_1^{(0)} = - 4 + \frac{8E_2}{M_1} - \frac{4M_2^2}{M_1^2} + \frac{16E}{M_1} \left[1 - \frac{E_2}{M_1}  - \frac{E}{M_1} \right] + \frac{m^2}{M_1^2} \left[ - 3 + \frac{6E_2}{M_1} + \frac{M_2^2}{M_1^2} + \frac{8E}{M_1} - \frac{m^2}{M_1^2} \right], \label{eq:a10}
\end{equation}
\begin{equation}
A_2^{(0)} = \frac{2m^2}{M_1^2} \left[3 - \frac{2E_2}{M_1} - \frac{M_2^2}{M_1^2} - \frac{4E}{M_1} + \frac{m^2}{M_1^2} \right], \label{eq:a20}
\end{equation}
and
\begin{equation}
A_3^{(0)} = \frac{m^2}{M_1^2} \left[ 1 - \frac{2E_2}{M_1} + \frac{M_2^2}{M_1^2} - \frac{m^2}{M_1^2} \right], \label{eq:a30}
\end{equation}
and the factor $d\Omega$ reads
\begin{equation}
d\Omega = C_K^2 \frac{G_F^2|V_{us}|^2}{32\pi^3}M_1^3 dEdE_2.
\end{equation}
A glance at expressions (\ref{eq:a10})-(\ref{eq:a30}) reveals why experiments about $K_{\mu 3}$ decays usually determine both $f_+$ and $f_-$, whereas experiments about $K_{e3}$ decays are only sensitive to $f_+$ because $f_-$ comes along with a term proportional to the positron (electron) mass squared, which renders its contribution negligible. In Tables \ref{t:eAi0} and \ref{t:mAi0} we present numerical evaluations of $A_i^{(0)}$ at various points $(E,E_2)$ of the Dalitz plots of $K_{e3}^\pm$ and $K_{\mu 3}^\pm$ decays, respectively, for the sake of completeness.

\begingroup
\squeezetable
\begin{table}
\caption{\label{t:eAi0}Values of $A_i^{(0)}$, Eqs.~(\ref{eq:a10})-(\ref{eq:a30}), in the TBR of the process $K^+ \to \pi^0 + e^+ + \nu_e$. The entries correspond to (a) $A_1^{(0)}$, (b) $A_2^{(0)}\times 10^5$, and (c) $A_3^{(0)}\times 10^6$. The energies $E$ and $E_2$ are given in $\textrm{GeV}$.}
\begin{ruledtabular}
\begin{tabular}{crrrrrrrrr}
$E_2\backslash E$ & $ 0.0123$ & $ 0.0370$ & $ 0.0617$ & $ 0.0864$ & $ 0.1111$ & $ 0.1358$ & $ 0.1604$ & $ 0.1851$ & $ 0.2098$ \\
\hline
 & & & & & (a) & & & \\
$0.2592$ & $ 0.0810$ & $ 0.3810$ & $ 0.6010$ & $ 0.7410$ & $ 0.8010$ & $ 0.7810$ & $ 0.6810$ & $ 0.5010$ & $ 0.2410$ \\
$0.2468$ &           & $ 0.2110$ & $ 0.4510$ & $ 0.6110$ & $ 0.6910$ & $ 0.6910$ & $ 0.6110$ & $ 0.4510$ & $ 0.2110$ \\
$0.2345$ &           & $ 0.0410$ & $ 0.3010$ & $ 0.4810$ & $ 0.5810$ & $ 0.6010$ & $ 0.5410$ & $ 0.4010$ & $ 0.1810$ \\
$0.2222$ &           &           & $ 0.1510$ & $ 0.3510$ & $ 0.4710$ & $ 0.5110$ & $ 0.4710$ & $ 0.3510$ & $ 0.1510$ \\
$0.2098$ &           &           & $ 0.0010$ & $ 0.2210$ & $ 0.3610$ & $ 0.4210$ & $ 0.4010$ & $ 0.3010$ & $ 0.1210$ \\
$0.1975$ &           &           &           & $ 0.0910$ & $ 0.2510$ & $ 0.3310$ & $ 0.3310$ & $ 0.2510$ & $ 0.0910$ \\
$0.1851$ &           &           &           &           & $ 0.1410$ & $ 0.2410$ & $ 0.2610$ & $ 0.2010$ & $ 0.0610$ \\
$0.1728$ &           &           &           &           & $ 0.0310$ & $ 0.1510$ & $ 0.1910$ & $ 0.1510$ & $ 0.0310$ \\
$0.1604$ &           &           &           &           &           & $ 0.0610$ & $ 0.1210$ & $ 0.1010$ & $ 0.0010$ \\
$0.1481$ &           &           &           &           &           &           & $ 0.0510$ & $ 0.0510$ &           \\
\hline
 & & & & & (b) & & & \\
$0.2592$ & $ 0.3804$ & $ 0.3375$ & $ 0.2947$ & $ 0.2518$ & $ 0.2090$ & $ 0.1661$ & $ 0.1233$ & $ 0.0804$ & $ 0.0376$ \\
$0.2468$ &           & $ 0.3483$ & $ 0.3054$ & $ 0.2625$ & $ 0.2197$ & $ 0.1768$ & $ 0.1340$ & $ 0.0911$ & $ 0.0483$ \\
$0.2345$ &           & $ 0.3590$ & $ 0.3161$ & $ 0.2733$ & $ 0.2304$ & $ 0.1875$ & $ 0.1447$ & $ 0.1018$ & $ 0.0590$ \\
$0.2222$ &           &           & $ 0.3268$ & $ 0.2840$ & $ 0.2411$ & $ 0.1983$ & $ 0.1554$ & $ 0.1126$ & $ 0.0697$ \\
$0.2098$ &           &           & $ 0.3375$ & $ 0.2947$ & $ 0.2518$ & $ 0.2090$ & $ 0.1661$ & $ 0.1233$ & $ 0.0804$ \\
$0.1975$ &           &           &           & $ 0.3054$ & $ 0.2625$ & $ 0.2197$ & $ 0.1768$ & $ 0.1340$ & $ 0.0911$ \\
$0.1851$ &           &           &           &           & $ 0.2733$ & $ 0.2304$ & $ 0.1875$ & $ 0.1447$ & $ 0.1018$ \\
$0.1728$ &           &           &           &           & $ 0.2840$ & $ 0.2411$ & $ 0.1983$ & $ 0.1554$ & $ 0.1126$ \\
$0.1604$ &           &           &           &           &           & $ 0.2518$ & $ 0.2090$ & $ 0.1661$ & $ 0.1233$ \\
$0.1481$ &           &           &           &           &           &           & $ 0.2197$ & $ 0.1768$ &           \\
\hline
 & & & & & (c) & & & \\
$0.2592$ & $ 0.0265$ & $ 0.0265$ & $ 0.0265$ & $ 0.0265$ & $ 0.0265$ & $ 0.0265$ & $ 0.0265$ & $ 0.0265$ & $ 0.0265$ \\
$0.2468$ &           & $ 0.0801$ & $ 0.0801$ & $ 0.0801$ & $ 0.0801$ & $ 0.0801$ & $ 0.0801$ & $ 0.0801$ & $ 0.0801$ \\
$0.2345$ &           & $ 0.1337$ & $ 0.1337$ & $ 0.1337$ & $ 0.1337$ & $ 0.1337$ & $ 0.1337$ & $ 0.1337$ & $ 0.1337$ \\
$0.2222$ &           &           & $ 0.1872$ & $ 0.1872$ & $ 0.1872$ & $ 0.1872$ & $ 0.1872$ & $ 0.1872$ & $ 0.1872$ \\
$0.2098$ &           &           & $ 0.2408$ & $ 0.2408$ & $ 0.2408$ & $ 0.2408$ & $ 0.2408$ & $ 0.2408$ & $ 0.2408$ \\
$0.1975$ &           &           &           & $ 0.2944$ & $ 0.2944$ & $ 0.2944$ & $ 0.2944$ & $ 0.2944$ & $ 0.2944$ \\
$0.1851$ &           &           &           &           & $ 0.3479$ & $ 0.3479$ & $ 0.3479$ & $ 0.3479$ & $ 0.3479$ \\
$0.1728$ &           &           &           &           & $ 0.4015$ & $ 0.4015$ & $ 0.4015$ & $ 0.4015$ & $ 0.4015$ \\
$0.1604$ &           &           &           &           &           & $ 0.4551$ & $ 0.4551$ & $ 0.4551$ & $ 0.4551$ \\
$0.1481$ &           &           &           &           &           &           & $ 0.5087$ & $ 0.5087$ &           \\
\end{tabular}
\end{ruledtabular}
\end{table}
\endgroup

\begingroup
\squeezetable
\begin{table}
\caption{\label{t:mAi0}Values of $A_i^{(0)}$, Eqs.~(\ref{eq:a10})-(\ref{eq:a30}), in the TBR of the process $K^+ \to \pi^0 + \mu^+ + \nu_\mu$. The entries correspond to (a) $A_1^{(0)}$, (b) $A_2^{(0)}$, and (c) $A_3^{(0)}\times 10$. The energies $E$ and $E_2$ are given in $\textrm{GeV}$.}
\begin{ruledtabular}
\begin{tabular}{crrrrrrrrr}
$E_2\backslash E$ & $ 0.1131$ & $ 0.1280$ & $ 0.1429$ & $ 0.1578$ & $ 0.1727$ & $ 0.1876$ & $ 0.2025$ & $ 0.2174$ & $ 0.2322$ \\
\hline
 & & & & & (a) & & & \\
$ 0.2480$ &           &           &           &           & $ 0.6777$ & $ 0.5767$ & $ 0.4466$ & $ 0.2874$ & $ 0.0990$ \\
$ 0.2361$ &           & $ 0.7065$ & $ 0.7045$ & $ 0.6734$ & $ 0.6131$ & $ 0.5238$ & $ 0.4053$ & $ 0.2577$ & $ 0.0810$ \\
$ 0.2242$ & $ 0.5682$ & $ 0.6070$ & $ 0.6166$ & $ 0.5972$ & $ 0.5486$ & $ 0.4708$ & $ 0.3640$ & $ 0.2280$ & $ 0.0629$ \\
$ 0.2123$ & $ 0.4571$ & $ 0.5075$ & $ 0.5288$ & $ 0.5210$ & $ 0.4840$ & $ 0.4179$ & $ 0.3227$ & $ 0.1984$ & $ 0.0449$ \\
$ 0.2004$ & $ 0.3460$ & $ 0.4080$ & $ 0.4410$ & $ 0.4447$ & $ 0.4194$ & $ 0.3650$ & $ 0.2814$ & $ 0.1687$ & $ 0.0269$ \\
$ 0.1885$ &           & $ 0.3086$ & $ 0.3531$ & $ 0.3685$ & $ 0.3549$ & $ 0.3120$ & $ 0.2401$ & $ 0.1390$ &           \\
$ 0.1766$ &           &           & $ 0.2653$ & $ 0.2923$ & $ 0.2903$ & $ 0.2591$ & $ 0.1988$ & $ 0.1094$ &           \\
$ 0.1647$ &           &           & $ 0.1774$ & $ 0.2161$ & $ 0.2257$ & $ 0.2062$ & $ 0.1575$ & $ 0.0797$ &           \\
$ 0.1528$ &           &           &           & $ 0.1399$ & $ 0.1612$ & $ 0.1532$ & $ 0.1162$ & $ 0.0501$ &           \\
$ 0.1409$ &           &           &           &           &           & $ 0.1003$ & $ 0.0749$ &           &           \\
\hline
 & & & & & (b) & & & \\
$ 0.2480$ &           &           &           &           & $ 0.0520$ & $ 0.0409$ & $ 0.0298$ & $ 0.0188$ & $ 0.0077$ \\
$ 0.2361$ &           & $ 0.0895$ & $ 0.0785$ & $ 0.0674$ & $ 0.0564$ & $ 0.0453$ & $ 0.0343$ & $ 0.0232$ & $ 0.0122$ \\
$ 0.2242$ & $ 0.1050$ & $ 0.0940$ & $ 0.0829$ & $ 0.0718$ & $ 0.0608$ & $ 0.0497$ & $ 0.0387$ & $ 0.0276$ & $ 0.0166$ \\
$ 0.2123$ & $ 0.1094$ & $ 0.0984$ & $ 0.0873$ & $ 0.0763$ & $ 0.0652$ & $ 0.0542$ & $ 0.0431$ & $ 0.0320$ & $ 0.0210$ \\
$ 0.2004$ & $ 0.1138$ & $ 0.1028$ & $ 0.0917$ & $ 0.0807$ & $ 0.0696$ & $ 0.0586$ & $ 0.0475$ & $ 0.0365$ & $ 0.0254$ \\
$ 0.1885$ &           & $ 0.1072$ & $ 0.0961$ & $ 0.0851$ & $ 0.0740$ & $ 0.0630$ & $ 0.0519$ & $ 0.0409$ &           \\
$ 0.1766$ &           &           & $ 0.1006$ & $ 0.0895$ & $ 0.0785$ & $ 0.0674$ & $ 0.0563$ & $ 0.0453$ &           \\
$ 0.1647$ &           &           & $ 0.1050$ & $ 0.0939$ & $ 0.0829$ & $ 0.0718$ & $ 0.0608$ & $ 0.0497$ &           \\
$ 0.1528$ &           &           &           & $ 0.0983$ & $ 0.0873$ & $ 0.0762$ & $ 0.0652$ & $ 0.0541$ &           \\
$ 0.1409$ &           &           &           &           &           & $ 0.0807$ & $ 0.0696$ &           &           \\
\hline
 & & & & & (c) & & & \\
$ 0.2480$ &           &           &           &           & $ 0.0110$ & $ 0.0110$ & $ 0.0110$ & $ 0.0110$ & $ 0.0110$ \\
$ 0.2361$ &           & $ 0.0331$ & $ 0.0331$ & $ 0.0331$ & $ 0.0331$ & $ 0.0331$ & $ 0.0331$ & $ 0.0331$ & $ 0.0331$ \\
$ 0.2242$ & $ 0.0552$ & $ 0.0552$ & $ 0.0552$ & $ 0.0552$ & $ 0.0552$ & $ 0.0552$ & $ 0.0552$ & $ 0.0552$ & $ 0.0552$ \\
$ 0.2123$ & $ 0.0773$ & $ 0.0773$ & $ 0.0773$ & $ 0.0773$ & $ 0.0773$ & $ 0.0773$ & $ 0.0773$ & $ 0.0773$ & $ 0.0773$ \\
$ 0.2004$ & $ 0.0994$ & $ 0.0994$ & $ 0.0994$ & $ 0.0994$ & $ 0.0994$ & $ 0.0994$ & $ 0.0994$ & $ 0.0994$ & $ 0.0994$ \\
$ 0.1885$ &           & $ 0.1215$ & $ 0.1215$ & $ 0.1215$ & $ 0.1215$ & $ 0.1215$ & $ 0.1215$ & $ 0.1215$ &           \\
$ 0.1766$ &           &           & $ 0.1435$ & $ 0.1435$ & $ 0.1435$ & $ 0.1435$ & $ 0.1435$ & $ 0.1435$ &           \\
$ 0.1647$ &           &           & $ 0.1656$ & $ 0.1656$ & $ 0.1656$ & $ 0.1656$ & $ 0.1656$ & $ 0.1656$ &           \\
$ 0.1528$ &           &           &           & $ 0.1877$ & $ 0.1877$ & $ 0.1877$ & $ 0.1877$ & $ 0.1877$ &           \\
$ 0.1409$ &           &           &           &           &           & $ 0.2098$ & $ 0.2098$ &           &           \\
\end{tabular}
\end{ruledtabular}
\end{table}
\endgroup

Now that we have some insight into the main features of kaon semileptonic decays, we proceed to the calculation of radiative corrections. Let us first discuss the virtual case and later the bremsstrahlung one.

\section{\label{sec:vir}Virtual radiative corrections}

\begin{figure}[ht]
\scalebox{0.4}{\includegraphics{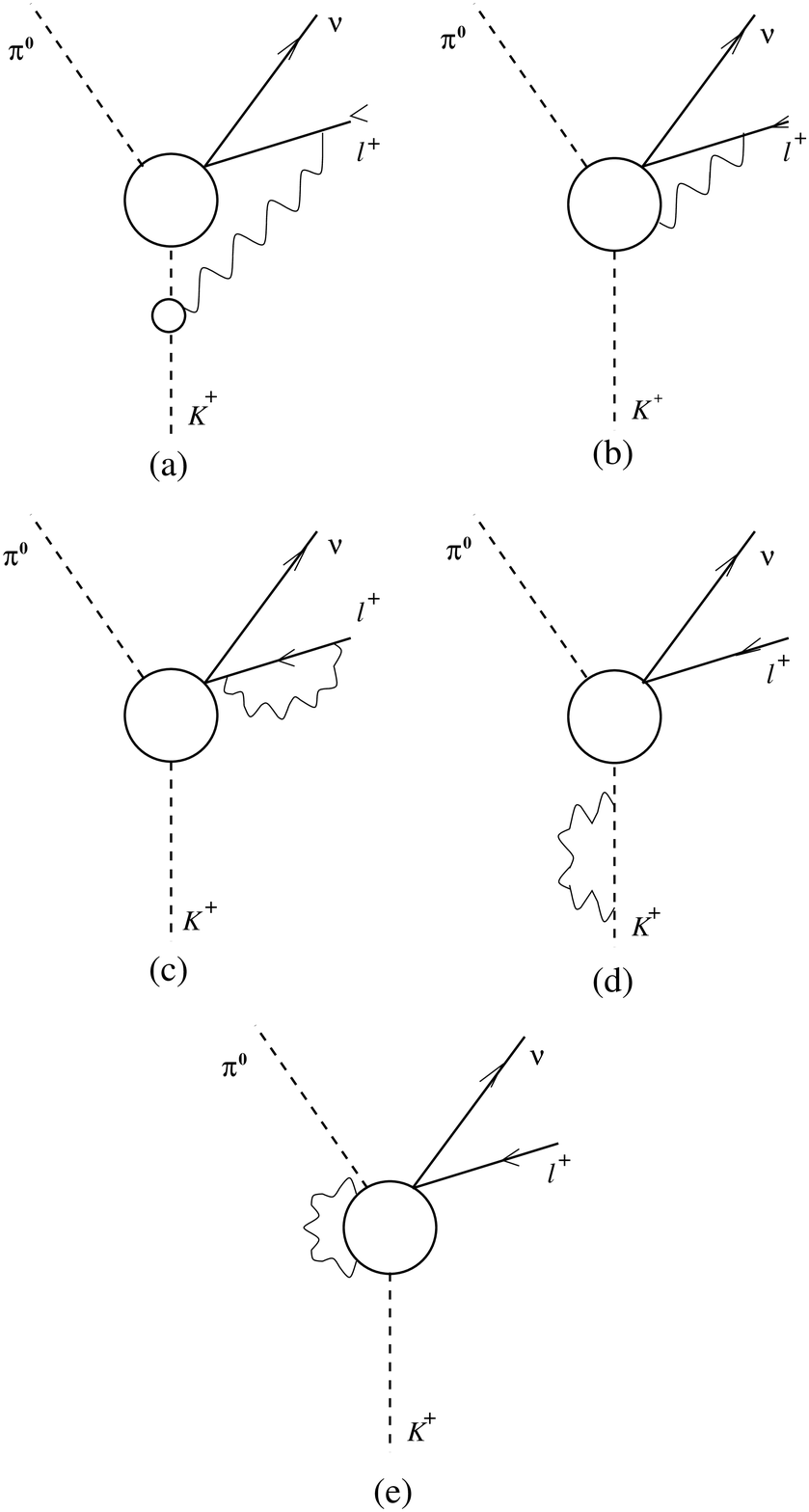}}
\caption{\label{fig:virtual}Feynman diagrams, to first order in $\alpha$, which yield virtual radiative corrections in $K_{l3}^+$ decay. The wavy, broken, and continuous lines represent virtual photons, pseudoscalar mesons, and fermions, respectively. The blobs represent the effects of the strong interactions, and, at the weak vertex, they also represent the effects of details of the weak interactions.}
\end{figure}

To first order in $\alpha$, the virtual radiative corrections in $K_{l3}^+$ decay arise from the analysis of the Feynman diagrams depicted in Fig.~\ref{fig:virtual}. These virtual radiative corrections can be computed in complete analogy with the approach implemented by Sirlin \cite{sirlin} in the study of the energy spectrum of the electron in neutron beta decay, subsequently extended to the analysis of hyperon semileptonic decays by Garc{\'\i}a and Ju\'arez \cite{juarez}. In this approach, the virtual radiative corrections can be separated in two parts. One part is model-independent and finite in the ultraviolet region and fully contains the infrared divergence. The other one depends on the details of the strong interactions. All in all, the separation relies on general principles such as Lorentz covariance, analyticity of the strong and weak interactions, and the validity of QED. Garc{\'\i}a and Maya \cite{maya} have already implemented the procedure to $M_{l3}^\pm$ decays, so we will partially borrow their methodology to achieve our goal. Further details on the procedure can be found in the original papers \cite{sirlin,juarez,maya}, so here we limit ourselves to describe only a few salient facts.

First of all, from the analysis of Fig.~\ref{fig:virtual} one can determine that Figs.~\ref{fig:virtual}(a)-\ref{fig:virtual}(b) comprise the graphs in which a photon is emitted from a hadronic line or the intermediate vector boson and is absorbed by the charged lepton. To the order of $(\alpha/\pi)(q/M_1)$, these diagrams yield the contribution \cite{maya}
\begin{equation}
\mathsf{M}_{V_1} = \frac{G}{\sqrt{2}} \frac{\alpha}{4\pi^3 i} \int d^{4}k \frac{D_{\mu\alpha}(k) \overline{u}_\nu O_\lambda (-2l_\alpha + \!\not{\!k}\gamma_\alpha)v_\ell}{k^2 - 2l\cdot k + i\varepsilon}
\left[ \frac{W_\lambda(p_1,p_2)(2{p_1}_\mu-k_\mu)}{k^2-2p_1 \cdot k + i\varepsilon} + T_{\mu\lambda} (p_1,p_2,k) \right], \label{eq:virtualM1}
\end{equation}
where $k$ and $D_{\mu\alpha}(k)$ are the photon four-momentum and propagator, respectively. The model dependence that arises from these diagrams is contained in $T_{\mu\lambda}$, whose explicit form is not needed here but can be obtained without difficulty by following Ref.~\cite{sirlin}. We only point out that $T_{\mu\lambda}$ is regular as $k\to 0$ and transverse in the sense that $k_\mu T_{\lambda\mu}=0$.

In the same way, the graph of Fig.~\ref{fig:virtual}(c) contains the positron wave function renormalization, and, after mass renormalization, it contributes \cite{maya}
\begin{equation}
\mathsf{M}_{V_2} = -\frac{G}{\sqrt{2}} \frac{\alpha}{8\pi^3 i}W_\lambda (p_1,p_2) \int d^4k D_{\mu\alpha}(k) \overline{u}_\nu O_\lambda \frac{(-\!\!\not{l}+m)}{2m^2} \frac{(2l_\alpha + \gamma_\alpha \!\not{\!k}) \!\!\not{l} \, (2l_\mu + \!\not{\!k} \gamma_\mu)}{(k^2 + 2l \cdot k+i\varepsilon)^2} v_\ell, \label{eq:virtualM2}
\end{equation}
which is infrared-divergent.

Finally, Figs.~\ref{fig:virtual}(d) and \ref{fig:virtual}(e) contain the graphs in which the photon is emitted by a hadronic line or the intermediate vector boson and is absorbed by the same hadronic line or another one or the intermediate boson. These figures thus yield \cite{maya}
\begin{eqnarray}
\mathsf{M}_{V_3} & = & \frac{G}{\sqrt{2}} \frac{\alpha}{8\pi^3 i} W_\lambda(p_1,p_2) \overline{u}_\nu O_\lambda v_\ell \int d^4k D_{\mu\alpha} (k) \frac{(2p_1-k)_\mu (2p_1-k)_\alpha}{(k^2-2p_1 \cdot k + i\varepsilon)^2} + \mathsf{M}_{V_3}^\prime \nonumber \\
& = & \mbox{} \mathsf{M}_{V_3}^c + \mathsf{M}_{V_3}^\prime, \label{eq:virtualM3}
\end{eqnarray}
where $\mathsf{M}_{V_3}^c$, explicitly defined in Eq.~(\ref{eq:virtualM3}), is infrared-divergent whereas $\mathsf{M}_{V_3}^\prime$, which can be written in the most general way as
\begin{equation}
\mathsf{M}_{V_3}^\prime = \frac{\alpha}{\pi} \frac{G}{\sqrt{2}} [a_1(q^2)(p_1+p_2)_\mu + a_2(q^2) (p_1-p_2)_\mu] \overline{u}_\nu \gamma_\mu(1+\gamma_5) v_\ell, \label{eq:mv3p}
\end{equation}
is infrared-convergent. The explicit expressions of the additional form factors $a_k(q^2)$ introduced in Eq.~(\ref{eq:mv3p}) are not needed for our purposes.

At this stage it is already possible to achieve the separation referred to above. The transition amplitude with virtual radiative corrections to the order of $(\alpha/\pi)(q/M_1)$ is constituted by two parts $\mathsf{M}_V^i$ and $\mathsf{M}_V^d$. The former is model-independent and gauge-invariant and contains in full the infrared divergence, whereas the latter contains all the model dependence. $\mathsf{M}_V^i$ is given by adding together the term proportional to the first summand within square brackets in Eq.~(\ref{eq:virtualM1}), $\mathsf{M}_{V_2}$, and $\mathsf{M}_{V_3}^c$. After some analysis we find
\begin{equation}
\mathsf{M}_V^i = \mathsf{M}_0 \frac{\alpha}{2\pi} \Phi_1(E) + \mathsf{M}_{p_1} \frac{\alpha}{2\pi} \Phi_2(E), \label{eq:MV}
\end{equation}
where the functions $\Phi_1(E)$ and $\Phi_2(E)$ are given by
\begin{eqnarray}
\Phi_1(E) & = & 2 \left[ \frac{1}{\beta} \mathrm{arctanh} \, \beta - 1 \right] \ln \left[ \frac{\lambda}{m} \right] -
\frac{1}{\beta}(\mathrm{arctanh} \, \beta)^2 + \frac{1}{\beta} L \left[\frac{b_1-b_2}{-b_2} \right] - \frac{1}{\beta} L \left[\frac{b_1-b_2}{1-b_2}\right] \nonumber \\
&  & \mbox{} + \frac{1}{\beta} \mathrm{arctanh} \, \beta
\left[ \frac{M_1^2-M_1E(1+\beta^2)}{b_3} \right] + \left[ \frac32-\frac{m^2}{b_3} \right] \ln \left[ \frac{M_{1}}{m}\right] \nonumber \\
&  & \mbox{} - \frac{1}{\beta} \ln \left[\frac{1-b_1}{1-b_2} \right] \left\{ \ln \left[ \frac{M_{1}}{m} \right] - \mathrm{arctanh} \, \beta \right\} - \frac{11}{8}, \label{eq:phiE}
\end{eqnarray}
and
\begin{equation}
\Phi_2(E) = \frac{1-\beta^2}{\beta} \left\{-\mathrm{arctanh} \, \beta \left[1+\frac{M_1E-m^2}{b_3} \right] +
\frac{M_1l}{b_3} \ln \left[\frac{M_1}{m}\right] \right\}, \label{eq:phiEp}
\end{equation}
where
\begin{subequations}
\begin{eqnarray}
b_{1,2} & = & \frac{m^2-M_1E \pm M_1l}{b_3}, \\
b_3 & = & M_1^2+m^2-2M_1E.
\end{eqnarray}
\end{subequations}
Here $\beta \equiv l/E$, $L$ is the Spence function, and $\lambda$ is the infrared-divergent cutoff. This divergent term
will be canceled by its counterpart in the bremsstrahlung contribution to be discussed in the next section.

Similarly, the second term in Eq.~(\ref{eq:MV}) is given by
\begin{equation}
\mathsf{M}_{p_1} = -\frac{E}{m M_1} \frac{G}{\sqrt 2} W_\lambda [\overline{u}_\nu O_\lambda \! \! \not \! p_1 v_\ell]. \label{eq:Mp1}
\end{equation}

As for the model-dependent part $\mathsf{M}_V^d$, it is given by the sum of the term proportional to the second summand within square brackets in Eq.~(\ref{eq:virtualM1}) and $\mathsf{M}_{V_3}^\prime$. The sheer impossibility of computing analytically the integrals over the photon four-momentum involved in $\mathsf{M}_V^d$ leads us to implement Lorentz invariance instead, so we get \cite{maya}
\begin{equation}
\mathsf{M}_V^d = \frac{\alpha}{\pi} \frac{G}{\sqrt{2}} [a_1^{\prime\prime} (q^2,p_+ \cdot l) (p_1+p_2)_\mu + a_2^{\prime\prime} (q^2,p_+ \cdot l) (p_1-p_2)_\mu ] \overline{u}_\nu \gamma_\mu (1+\gamma_5) v_\ell,
\end{equation}
where $a_1^{\prime \prime}$ and $a_2^{\prime \prime}$ are some other form factors which now depend on $q^2$ and $p_+\cdot l = (p_1+p_2) \cdot l$. A close inspection of $\mathsf{M}_V^d$ reveals that it has the same structure as $\mathsf{M}_0$ [Eq.~(\ref{eq:M0})], so one is prompted to absorb the former into the latter by redefining the original form factors $f_\pm$, namely,
\begin{eqnarray}
\mathsf{M}_0^\prime & = & \mathsf{M}_0 + \mathsf{M}_V^d \nonumber \\
& = & \mbox{} \frac{G}{\sqrt 2} [f_+^\prime(q^2,p_+\cdot l)(p_1+p_2)_\mu + f_-^\prime(q^2,p_+\cdot l)(p_1-p_2)_\mu] \overline{u}_\nu(p_\nu) O_\mu v_\ell(l), \label{eq:M0prime}
\end{eqnarray}
where the modified form factors $f_+^\prime$ and $f_-^\prime$ have a new dependence in the positron and emitted pion energies other than the ones in the $q^2$ dependence of the original form factors. By rearranging some terms of the order of $\mathcal{O}(\alpha)$, these form factors can be written generically as
\begin{subequations}
\label{eq:mff}
\begin{eqnarray}
f_+^\prime (q^2,p_+\cdot l) & = & f_+ (q^2) + \frac{\alpha}{\pi} a_+(p_+\cdot l), \\
f_-^\prime (q^2,p_+\cdot l) & = & f_- (q^2) + \frac{\alpha}{\pi} a_-(p_+\cdot l),
\end{eqnarray}
\end{subequations}
where $a_+$ and $a_-$ are some other functions which contain all the model dependence and the prime on $\mathsf{M}_0$ in Eq.~(\ref{eq:M0prime}) will be used as a reminder of this fact.

Let us remark that the introduction of the modified form factors $f_\pm^\prime(q^2,p_+\cdot l)$ in Eq.~(\ref{eq:M0prime}) is not a withdrawal of the approach. Ultimately, these modified form factors are the ones which can be experimentally measured and are the ones that provide information about the strong interactions.

By gathering together partial results, the transition amplitude for $K_{l3}^+$ decay with virtual radiative corrections is then given by
\begin{equation}
\mathsf{M}_V = \mathsf{M}_0^\prime \left[1+\frac{\alpha}{2\pi} \Phi_1(E) \right] + \mathsf{M}_{p_1} \frac{\alpha}{2\pi} \Phi_2(E), \label{eq:MVir}
\end{equation}
where some terms of second order in $\alpha$ have been rearranged in the above equation to stress the fact that to first order in $\alpha$ only the modified form factors $f_\pm(q^2,p_+\cdot l)$ appear in it.

Armed with the transition amplitude with virtual radiative corrections, $\mathsf{M}_V$, we can obtain the corresponding differential decay rate $d\Gamma_V$ by means of a long and tedious but otherwise standard procedure. Assuming for definiteness complex form factors, the final expression can be cast into
\begin{equation}
d\Gamma_V = d\Omega \left[A_0^\prime \left(1+\frac{\alpha}{\pi} \Phi_1\right) + A_V^\prime \left(\frac{\alpha}{\pi} \Phi_2 \right) \right], \label{eq:dGVfinal}
\end{equation}
where
\begin{equation}
A_V^\prime = A_1^{(V)} |f_+^\prime(q^2,p_+\cdot l)|^2 + A_2^{(V)}\textrm{Re} \,[f_+^\prime (q^2,p_+\cdot l){f_-^{\prime}}^*(q^2,p_+\cdot l)] + A_3^{(V)} |f_-^\prime(q^2,p_+\cdot l)|^2,  \label{eq:a0p}
\end{equation}
with
\begin{equation}
A_1^{(V)} = \frac{4E}{M_1} \left[ 1-\frac{M_2^2}{M_1^2} - \frac{E}{M_1} \left(\frac32 + \frac{E_2}{M_1} - \frac{M_2^2}{2M_1^2} \right) + \frac{m^2}{2M_1^2} \left( 1 + \frac{E_2}{M_1}\right) \right],
\end{equation}
\begin{equation}
A_2^{(V)}=\frac{4E}{M_1} \left[ \left( 1+\frac{M_2^2}{M_1^2} - \frac{2E_2}{M_1} \right) \left(1 - \frac{E}{M_1} \right) -\frac{m^2E_2}{M_1^3}\right],
\end{equation}
and
\begin{equation}
A_3^{(V)} = \frac{E^2}{2M_1^2} \left[ 1-\frac{2E_2}{M_1} + \frac{M_2^2}{M_1^2} - \frac{m^2 \left( M_1 - E_2 \right)}{M_1^2E} \right] .
\end{equation}

Equation~(\ref{eq:dGVfinal}) is our first partial result for $K_{l3}^\pm$ decays. It is the differential decay rate with virtual radiative corrections to the order of $(\alpha/\pi)(q/M_1)$. It contains an infrared-divergent term in $\Phi_1(E)$ which at any rate will be canceled when the bremsstrahlung radiative corrections are added. Let us stress the fact that all the model dependence arising from the virtual contribution has been absorbed into the modified form factors (\ref{eq:mff}) which are the ones that enter into Eq.~(\ref{eq:dGVfinal}). The primes in this equation are an indicator of this fact. Let us now discuss the bremsstrahlung contribution.

\section{\label{sec:bremss}Bremsstrahlung radiative corrections}

In this section we now turn to the analysis of the emission of a real photon in the process
\begin{equation}
K^+(p_1) \to \pi^0(p_2) + \ell^+(l) + \nu_\ell(p_\nu) + \gamma(k), \label{eq:kl3gamma}
\end{equation}
where $K^+$ denotes a positively charged kaon and $\pi^0$ a neutral pion, whereas $\ell$ stands for a positively charged lepton ($\ell = e^+$ or $\mu^+$) and $\nu_\ell$ its accompanying neutrino. We again point out that the charge conjugate mode of process (\ref{eq:kl3gamma}) can be analyzed likewise. Here, $\gamma$ represents a real photon with four-momentum $k=(\omega,\mathbf{k})$. As in the case of virtual radiative corrections, we will use the rest system of the decaying kaon. Accordingly, energy and momentum conservation yield $M_1 = E+E_2+E_\nu+\omega$ and $\mathbf{0} = \mathbf{p}_2 + \mathbf{l} + \mathbf{p}_\nu + \mathbf{k}$, where the neutrino energy and momentum in the presence of the photon are, respectively,
\begin{equation}
E_\nu = E_\nu^0-\omega, \qquad \quad \mathbf{p}_\nu = \mathbf{p}_\nu^0 - \mathbf{k}, \label{eq:Enu}
\end{equation}
where $E_\nu^0$ and $\mathbf{p}_\nu^0$ are the energy and three-momentum of the neutrino in the nonradiative process (\ref{eq:kl3p}), respectively.

A quick glance at Fig.~\ref{fig:kinem} reveals that the TBR of the Dalitz plot is the region where the three-body decay (\ref{eq:kl3p}) and the four-body decay (\ref{eq:kl3gamma}) overlap completely. In contrast, the FBR is where in process (\ref{eq:kl3gamma}) neither of the energies of the neutrino and the real photon can be zero. Therefore, strictly speaking, process (\ref{eq:kl3gamma}) is kinematically allowed to occur anywhere in the joined area $\textrm{TBR}\cup \textrm{FBR}$ of Fig.~\ref{fig:kinem}. If we assume that real photons can be discriminated in an experimental setup, then our analysis of bremsstrahlung radiative corrections will consider process (\ref{eq:kl3gamma}) restricted to the TBR.

\begin{figure}[ht]
\scalebox{0.4}{\includegraphics{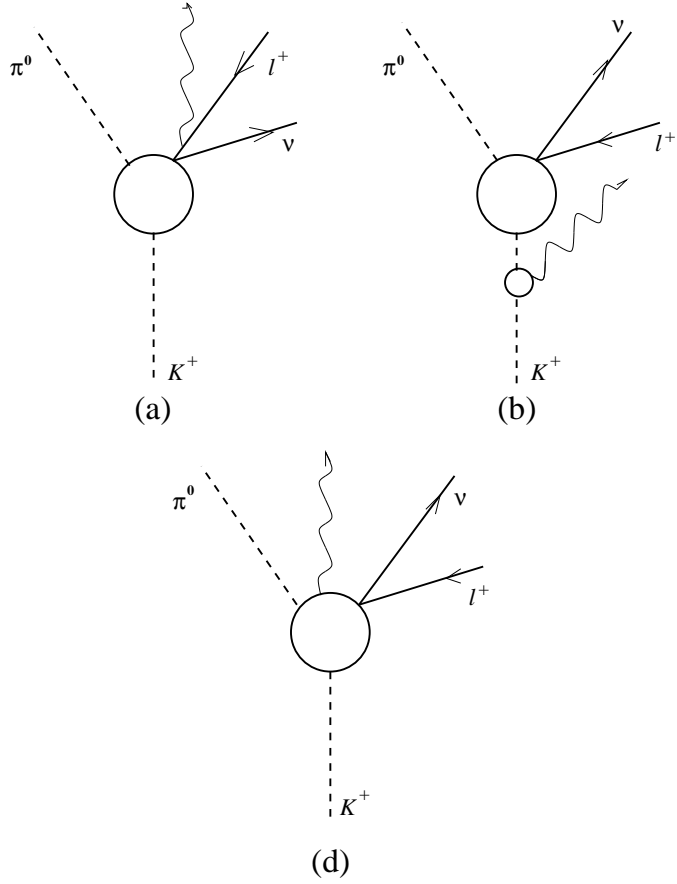}}
\caption{\label{fig:bremss}Feynman diagrams which yield bremsstrahlung radiative corrections in $K_{l3}^+$ decay. The wavy lines represent real photons, the broken lines represent pseudoscalar mesons, and the continuous lines represent fermions.}
\end{figure}

To first order in $\alpha$, the bremsstrahlung radiative corrections in $K_{l3}^\pm$ decays arise from the analysis of the Feynman diagrams depicted in Fig.~\ref{fig:bremss}. We now need to obtain the transition amplitude of process (\ref{eq:kl3gamma}), $\mathsf{M}_B$, keeping all the terms of the order of $(\alpha/\pi)(q/M_1)$ explicitly. This can be achieved in a model-independent way by using the Low theorem \cite{low,chew}, which asserts that the radiative amplitudes of orders $1/k$ and $(k)^0$ can be determined in terms of the nonradiative amplitude without further structure dependence. Therefore, the direct application of the Low theorem allows us to express $\mathsf{M}_B$ as
\begin{equation}
\mathsf{M}_B =\sum_{i=1}^{4} \mathsf{M}_{B_i}, \label{eq:mb}
\end{equation}
where
\begin{equation}
\mathsf{M}_{B_1} = -\frac{eG}{\sqrt{2}} W_{\lambda} [\overline{u}_\nu O_\lambda v_\ell]  \left[ \frac{l\cdot \epsilon}{l\cdot k} - \frac{p_1 \cdot \epsilon}{p_1\cdot k} \right], \label{eq:mb1}
\end{equation}
\begin{equation}
\mathsf{M}_{B_2} = -\frac{eG}{\sqrt{2}} W_{\lambda} \overline{u}_\nu O_\lambda \frac{\not{\!k} \!\! \not{\!\epsilon}}{2l \cdot k} v_\ell, \label{eq:mb2}
\end{equation}
\begin{equation}
\mathsf{M}_{B_3} = -\frac{eG}{\sqrt{2}} (f_+ + f_-) \left[ \frac{p_1 \cdot \epsilon}{p_1 \cdot k} k_\lambda - \epsilon_\lambda \right] \overline{u}_\nu O_\lambda v_\ell, \label{eq:mb3}
\end{equation}
and
\begin{equation}
\mathsf{M}_{B_4} = -\frac{eG}{\sqrt{2}} \left[ \frac{p_1 \cdot \epsilon}{p_1 \cdot k} q\cdot k-q\cdot \epsilon \right] \frac{\partial W_\lambda}{\partial q^2} \, \overline{u}_\nu O_\lambda v_\ell. \label{eq:mb4}
\end{equation}
Here $\epsilon$ is the polarization four-vector of the photon. Let us notice that, strictly speaking, $\mathsf{M}_{B_4}$ in Eq.~(\ref{eq:mb4}) will contribute to the order of $\mathcal{O}(q^2)$ to the decay rate so it will be suppressed with respect to the others and can be ignored in our analysis.

Now, the differential decay rate with bremsstrahlung radiative corrections $d\Gamma_B$ can be written as \cite{bd}
\begin{equation}
d\Gamma_B = \frac{1}{(2\pi)^8} \frac{1}{2M_1}\frac{mm_\nu}{4E_2EE_\nu\omega} d^3p_2\,d^3l\,d^3p_\nu\,d^3k \, \delta^4(p_1-p_2-l-p_\nu-k) \sum_{\textrm{spins},\epsilon}|\mathsf{M}_B|^2. \label{eq:diffdgb}
\end{equation}

In order to correctly account for the contribution of the unobserved photons to the Dalitz plot of process (\ref{eq:kl3gamma}) restricted to the TBR, we need to perform a careful analysis to delimit the integrations over the kinematical variables in (\ref{eq:diffdgb}). We have already mentioned that the system of reference we use is the rest frame of the decaying kaon. Accordingly, with all generality, we use the same orientation of the coordinate axes as in the previous cases: namely, the direction of emission of $\ell^+$ coincides with the $+z$ axis and $\pi^0$ is emitted in the first or fourth quadrant of the $(x,z)$ plane. With this choice we are left with five out of 12 variables of the final state. Two of them should be, of course, the energies $E$ and $E_2$ of $\ell^+$ and $\pi^0$, respectively, whose limits are given in Eq.~(\ref{eq:EE2lim}). The other three nontrivial variables can be grouped into two sets, namely, $(k,\cos\theta_k,\varphi_k)$ and $(\cos\theta_2,\cos\theta_k,\varphi_k)$, where $k$, $\theta_k$, and $\varphi_k$ are the magnitude of the three-momentum and the polar and azimuthal angles of the photon, respectively, and $\theta_2$ is the polar angle of $\pi^0$. The former set is more suitable for dealing with the infrared-divergent terms, whereas the latter is more useful in the analysis of the infrared-convergent contributions. For this latter case one has $\hat{\mathbf{p}}_2\cdot\hat{\mathbf{l}} = \cos\theta_2 \equiv y$, $\hat{\mathbf{l}} \cdot \hat{\mathbf{k}} = \cos\theta_k \equiv x$, and $\hat{\mathbf{p}}_2\cdot\hat{\mathbf{k}} = \cos\theta_2\cos\theta_k+\sin\theta_2\sin\theta_k\cos\varphi_k$, and the photon energy can be expressed as
\begin{equation}
\omega = \frac{F}{2D}, \label{eq:defw}
\end{equation}
where
\begin{subequations}
\begin{eqnarray}
F & = & 2p_2l(y_0-y), \\
D & = & E_\nu^0 + lx+\mathbf{p}_2\cdot\hat{\mathbf{k}},
\end{eqnarray}
\end{subequations}
and the scalar product $\hat{\mathbf{p}}_2\cdot\hat{\mathbf{k}}$ can readily be written in terms of $x$ and $y$.

As for $d\Gamma_B$ we consider it convenient to organize it as
\begin{equation}
d\Gamma_B=d\Gamma_{B_1} + d\Gamma_{B_2} + d\Gamma_{B_3} + d\Gamma_{B_4} + d\Gamma_{B_5}, \label{eq:dGB}
\end{equation}
where $d\Gamma_{B_1} \sim \sum |\mathsf{M}_{B_1}|^{2}$, $d\Gamma_{B_2} \sim \sum \left[ |\mathsf{M}_{B_2}|^2 + 2\mathrm{Re}[\mathsf{M}_{B_1} \mathsf{M}_{B_2}^\dagger]\right]$, $d\Gamma_{B_3} \sim \sum 2\mathrm{Re}[\mathsf{M}_{B_1} \mathsf{M}_{B_3}^\dagger]$, $d\Gamma_{B_4} \sim \sum 2\mathrm{Re}[\mathsf{M}_{B_2}\mathsf{M}_{B_3}^\dagger]$, and
$d\Gamma_{B_5} \sim \sum |\mathsf{M}_{B_3}|^2$.

The contribution $d\Gamma_{B_1}$ deserves particular attention because it not only contains an infrared-divergent term but also finite ones that come along with it, which must be properly identified. For this purpose, let us notice that $\sum|\mathsf{M}_{B_1}|^2$ can be conveniently separated as
\begin{equation}
\sum_{\mathrm{spins,\epsilon}} |\mathsf{M}_{B_1}|^2 = \frac{e^2G^2}{2} \frac{M_1^4}{m m_\nu} \left(A_0 + B\right) \sum_\epsilon \left[\frac{l\cdot \epsilon}{l\cdot k} - \frac{p_1\cdot\epsilon}{p_1\cdot k} \right]^2, \label{eq:MB12}
\end{equation}
where $l$, $p_1$, $k$, and $\epsilon$ within the square brackets in Eq.~(\ref{eq:MB12}) are understood to be four-vectors. Besides, $A_0$ is given in Eq.~(\ref{eq:a0}), and the function $B$ reads
\begin{equation}
B = B_1 |f_+(q^2)|^2 + B_2 \mathrm{Re} \,[f_+(q^2) f_-^*(q^2)] + B_3 |f_-(q^2)|^2, \label{eq:bb}
\end{equation}
where
\begin{eqnarray}
B_1 & = & -\frac{F^2}{M_1^4} + \frac{F}{M_1^2} \left[5-\frac{2E_2}{M_1}+\frac{M_2^2}{M_1^2}-\frac{8E}{M_1}+\frac{2m^2}{M_1^2}-\frac{2\omega}{M_1}-\frac{2\omega(E_2-\mathbf{p}_2\cdot\hat{\mathbf{k}})}{M_1^2} \right] \nonumber \\
&  & \mbox{} + \frac{2E\omega(1-\beta x)}{M_1^2}\left[1+\frac{2E_2}{M_1}+\frac{M_2^2}{M_1^2}\right] \nonumber \\
&  & \mbox{} + \frac{2\omega(M_1+E_2-\mathbf{p}_2\cdot\hat{\mathbf{k}})}{M_1^2} \left[1-\frac{2E_2}{M_1}+\frac{M_2^2}{M_1^2}-\frac{4E}{M_1}+\frac{m^2}{M_1^2}\right], \label{eq:b1}
\end{eqnarray}
\begin{eqnarray}
B_2 & = & \frac{2F}{M_1^2} \left[-\frac{2 E\omega (1-\beta x)}{M_1^2} - \frac{m^2}{M_1^2} + \frac{2\omega}{M_1} \right] + \frac{4E\omega(1-\beta x)}{M_1^2}\left[2-\frac{2E_2}{M_1}-\frac{4E}{M_1}+\frac{m^2}{M_1^2} \right] \nonumber \\
&  & \mbox{} + \frac{4\omega}{M_1}\left[-1+\frac{2E_2}{M_1}-\frac{M_2^2}{M_1^2}+\frac{2E}{M_1}-\frac{m^2}{M_1^2} \right] + \frac{\omega(E_2-\mathbf{p}_2\cdot\hat{\mathbf{k}})}{M_1^2} \left[-\frac{8E}{M_1} \right], \label{eq:b2}
\end{eqnarray}
and
\begin{equation}
B_3 = \frac{F}{M_1^2}\frac{m^2}{M_1^2} + \frac{2E\omega(1-\beta x)}{M_1^2} \left[\frac{F}{M_1^2}-\frac{m^2}{M_1^2} \right], \label{eq:b3}
\end{equation}

Equation (\ref{eq:MB12}) has been purposely separated the way it stands because the factor $A_0$ is precisely the one that is needed to cancel the infrared divergence contained in its virtual counterpart Eq.~(\ref{eq:dGVfinal}). Now, the extraction of the infrared divergence and the finite terms that come along with it can be conveniently performed by following either one of the approaches implemented by Ginsberg \cite{ginsKl3pm} (introduced in the analysis of $K_{l3}^\pm$ decays) or by Tun \textit{et.\ al.}\ \cite{tun89} (implemented in the analysis of hyperon semileptonic decays). Ultimately, it has been shown that both approaches are equivalent and yield the same results \cite{rfm97}. For convenience, in this work we follow the latter approach, and in Appendix \ref{app:drbr} we briefly describe some important aspects of the procedure.

On the other hand, we should exercise some caution when performing the pending sum over the photon polarization in Eq.~(\ref{eq:mb1}), which also raises an important issue. In all infrared-convergent terms the ordinary covariant summation can be used, namely, $\sum(\epsilon \cdot u) (\epsilon \cdot v) = -u\cdot v$, where $u=(u_0,\mathbf{u})$ and $v=(v_0,\mathbf{v})$ are arbitrary four-vectors and $\omega=k$, with $k$ the magnitude of $\mathbf{k}$. However, in the infrared-divergent terms the longitudinal degree of polarization of the photon must be accounted for. This can be carried out by using the Coester representation \cite{coester} in which
\begin{equation}
\sum_\epsilon (\epsilon \cdot u)(\epsilon \cdot v) = \mathbf{u} \cdot \mathbf{v} - \frac{(\mathbf{u} \cdot \mathbf{k}) (\mathbf{v} \cdot \mathbf{k})}{\omega^2}, \label{eq:coester}
\end{equation}
where $\omega^2=k^2+\lambda^2$ and $\lambda$ is a fictitious mass given to the photon to regularize the infrared divergence.

By taking into account all the above arguments, $d\Gamma_{B_1}$ can be written as
\begin{equation}
d\Gamma_{B_1} = d\Gamma_{B_1}^{\mathrm{ir}} + d\Gamma_{B_1}^{\mathrm{ic}},
\end{equation}
where $d\Gamma_{B_1}^{\mathrm{ir}}$, the piece containing the infrared divergence, can be evaluated through a direct application of Eq.~(\ref{eq:dB1}); the resultant expression is
\begin{eqnarray}
d\Gamma_{B_1}^{\mathrm{ir}} & = & \frac{\alpha}{\pi} d\Omega \frac{p_2l}{2\pi}\beta^2 A_0 \int_{-1}^1 dx \bigg[  \int_0^{2\pi}d\varphi_k\int_0^{k_4} dk\frac{k^2}{\omega} |g^-(\theta_2)| \frac{1-k^2x^2/\omega^2}{(\omega-\beta k x)^2} \nonumber \\
&  & \mbox{\hglue1.0truecm} + \int_{\pi/2}^{3\pi/2}d\varphi_k\int_{k_4}^{k_2} dk\frac{k^2}{\omega} |g^-(\theta_2)| \frac{1-k^2x^2/\omega^2}{(\omega-\beta k x)^2} \nonumber \\
&  & \mbox{\hglue1.0truecm} + \int_{\pi/2}^{3\pi/2}d\varphi_k\int_{k_4}^{k_2} dk\frac{k^2}{\omega} |g^+(\theta_2)| \frac{1-k^2x^2/\omega^2}{(\omega-\beta k x)^2} \bigg], \label{eq:dB1AA}
\end{eqnarray}
where we have used the definition \cite{tun89}
\begin{equation}
g^\pm(\theta_2) = \frac{\sin\theta_2^\pm}{a\sin\theta_2^\pm-b\cos\theta_2^\pm}, \label{eq:gth}
\end{equation}
and the factors that appear in Eqs.~(\ref{eq:dB1AA}) and (\ref{eq:gth}) are defined in Appendix \ref{app:drbr}.

Following the lines of Ref.~\cite{tun89}, we can split the range of integration over $k$ into two intervals, namely $(0,\Delta k)$ and $(\Delta k,k_i)$, with $i=2,4$. This approach is general enough to avoid taking the limit $\Delta k\to 0$ in the infrared-divergent terms because $\Delta k$ cancels exactly before performing the $\varphi_k$ integration. In contrast, the limit $\Delta k\to 0$ can be taken in the infrared-convergent terms because they are regular in $\Delta k$. Let us notice that in the interval $0\leq k \leq \Delta k$, the function $g^-(\theta_2)$ can be expanded in powers of $k$; the expansion yields \cite{tun89}
\begin{equation}
g^-(\theta_2) = \frac{1}{2p_2l}+\mathcal{O}(k).
\end{equation}
Thus, after some rearrangements, $d\Gamma_{B_1}^{\mathrm{ir}}$ becomes
\begin{eqnarray}
d\Gamma_{B_1}^{\mathrm{ir}} & = & \frac{\alpha}{\pi} d\Omega \frac{p_2l}{2\pi}\beta^2 A_0 \int_{-1}^1 dx \bigg[ \frac{1}{2p_2l} \int_0^{2\pi}d\varphi_k\int_0^{\Delta k} dk\frac{k^2}{\omega} \frac{1-k^2x^2/\omega^2}{(\omega-\beta k x)^2} \nonumber \\
&  & \mbox{\hglue1.3truecm} + \int_{-\pi/2}^{\pi/2}d\varphi_k\int_{\Delta k}^{k_4} dk\frac{k^2}{\omega} |g^-(\theta_2)| \frac{1-k^2x^2/\omega^2}{(\omega-\beta k x)^2} \nonumber\\
&  & \mbox{\hglue1.3truecm} + \int_{\pi/2}^{3\pi/2}d\varphi_k\int_{\Delta k}^{k_2} dk\frac{k^2}{\omega} |g^-(\theta_2)| \frac{1-k^2x^2/\omega^2}{(\omega-\beta k x)^2} \nonumber\\
&  & \mbox{\hglue1.3truecm} + \int_{\pi/2}^{3\pi/2}d\varphi_k\int_{k_4}^{k_2} dk\frac{k^2}{\omega} |g^+(\theta_2)| \frac{1-k^2x^2/\omega^2}{(\omega-\beta k x)^2} + \mathcal{O}(\Delta k) \bigg], \label{eq:dB1F1}
\end{eqnarray}

On the other hand, $y$ can also be expanded in powers of $k$ \cite{tun89}, namely,
\begin{equation}
y = y_0 - f^\prime k + \mathcal{O}(k^2),
\end{equation}
where
\begin{equation}
f^\prime = \frac{1}{p_2l} \left[E_\nu^0+(p_2y_0+l)x+p_2(1-y_0^2)^{1/2}(1-x^2)^{1/2}\cos\varphi_k \right].
\end{equation}
Thus, the last three summands in Eq.~(\ref{eq:dB1F1}) can be cast into the form (\ref{eq:dB2}), replacing the upper limit over the $y$ integration with $y_0-f^\prime \Delta k$, and also by using the definition of the photon energy, Eq.~(\ref{eq:defw}), namely,
\begin{eqnarray}
d\Gamma_{B_1}^{\mathrm{ir}} & = & \frac{\alpha}{\pi} d\Omega \frac{p_2l}{2\pi}\beta^2 A_0 \bigg[
\frac{1}{2p_2l} \int_0^{2\pi}d\varphi_k \int_{-1}^1 dx \int_0^{\Delta k} dk\frac{k^2}{\omega} \frac{1-k^2x^2/\omega^2}{(\omega-\beta k x)^2} \nonumber \\
&  & \mbox{\hglue1.3truecm} + \frac{1}{2p_2l} \int_{-1}^1 dx\frac{1-x^2}{(1-\beta x)^2} \int_{0}^{2\pi}d\varphi_k\int_{-1}^{y_0-f^\prime \Delta k} dy\frac{1}{y_0-y} \bigg]. \label{eq:dB1F3}
\end{eqnarray}
Now, the $\varphi_k$ integration in the first summand within the square brackets in Eq.~(\ref{eq:dB1F3}) can be trivially performed, whereas in the second summand we have rewritten the integral with the $x$ integration outermost so we can easily compute the $y$ integration. This results in
\begin{eqnarray}
d\Gamma_{B_1}^{\mathrm{ir}} & = & \frac{\alpha}{\pi} d\Omega A_0 \bigg[
\frac{\beta^2}{2} \int_{-1}^1 dx \int_0^{\Delta k} dk\frac{k^2}{\omega} \frac{1-k^2x^2/\omega^2}{(\omega-\beta k x)^2} \nonumber \\
&  & \mbox{\hglue1.3truecm} -\frac{\beta^2}{4\pi} \int_{-1}^1 dx \frac{1-x^2}{(1-\beta x)^2} \int_{0}^{2\pi} d\varphi_k \ln\left[\frac{f^\prime \Delta k}{1+y_0}\right] \bigg], \label{eq:dB1F4}
\end{eqnarray}
Finally, the two integrals in Eq.~(\ref{eq:dB1F4}) make up the function $\theta_1$ defined in Eq.~(96) of Ref.~\cite{tun89}, which contains an infrared-divergent piece with the correct coefficient to cancel its counterpart in the virtual radiative corrections. With this last result, we have achieved the proper identification of the infrared divergence and all the finite terms that come along with it.

We have already pointed out that the approach introduced in Ref.~\cite{ginsKe3pmDP} to deal with the infrared divergence is equivalent to the one described here. The equivalence is established through the relation $\theta_1 = I_0(E,E_2)$, where $I_0(E,E_2)$, given in Eq.~(27) of Ref.~\cite{ginsKe3pmDP}, in our notation reads
\begin{eqnarray}
I_0(E,E_2) & = & \frac{1}{\beta} \mathrm{arctanh} \beta \left[2 \ln\left(\frac{2l}{\lambda} \right) + \ln\left(\frac{m\eta_m^2}{4(E+l)r_+}\right)\right] - \frac{1}{\beta} L\left[-\frac{t^2}{4r_+} \right]
+ \frac{1}{\beta} L\left[-\frac{4r_-}{t^2} \right] \nonumber \\
&  & \mbox{} -2\ln\left[\frac{m}{\lambda} \right] - \ln \left[\frac{\eta_m^2}{2mE_\nu^0(q^2-m^2)} \right], \label{eq:i0}
\end{eqnarray}
where
\begin{equation}
(E+l)r_\pm = \left[E_\nu^0l^2(q^2-m^2) - \frac14 t^2E \right] \pm \left\{ \left[E_\nu^0l^2(q^2-m^2) - \frac14 t^2E\right]^2 - \frac{1}{16} m^2t^4 \right\}^{1/2},
\end{equation}
and
\begin{equation}
t^2=\eta_m(4p_2l-\eta_m), \qquad \eta_m=2p_2l(1+y_0).
\end{equation}

In summary, the final form of $d\Gamma_{B_1}^{\textrm{ir}}$ reduces to
\begin{equation}
d\Gamma_{B_1}^{\textrm{ir}} = \frac{\alpha}{\pi}d\Omega A_0 \theta_1. \label{eq:dgbir}
\end{equation}

As for the infrared-convergent contributions, the use of the set of variables $(\cos\theta_2,\cos\theta_k,\varphi_k)$, corresponding to the form (\ref{eq:dB2}), allows us to discern a group of triple integrals which make up $d\Gamma_{B_1}^{\textrm{ic}}$, $d\Gamma_{B_2},\ldots,d\Gamma_{B_5}$. Skipping details, these contributions read
\begin{equation}
d\Gamma_{B_1}^{\textrm{ic}} = \frac{\alpha}{\pi}d\Omega \frac{8}{M_1^2} \left[(\Lambda_1+\Lambda_2+\Lambda_3) |f_+(q^2)|^2 + (-\Lambda_2-2\Lambda_3) \mathrm{Re} \,[f_+(q^2) f_-^*(q^2)] + \Lambda_3 |f_-(q^2)|^2 \right], \label{eq:dgbic}
\end{equation}
\begin{equation}
d\Gamma_{B_2} = \frac{\alpha}{\pi}d\Omega \frac{8}{M_1^2} \left[(\Lambda_4+\Lambda_5+\Lambda_6) |f_+(q^2)|^2 + (-\Lambda_5-2\Lambda_6) \mathrm{Re} \,[f_+(q^2) f_-^*(q^2)] + \Lambda_6 |f_-(q^2)|^2 \right], \label{eq:dgb2}
\end{equation}
\begin{equation}
d\Gamma_{B_3} = \frac{\alpha}{\pi}d\Omega \frac{8}{M_1^2} \left[(\Lambda_7+\Lambda_8) |f_+(q^2)|^2 + \Lambda_7 \mathrm{Re} \,[f_+(q^2) f_-^*(q^2)] - \Lambda_8 |f_-(q^2)|^2 \right], \label{eq:dgb3}
\end{equation}
\begin{equation}
d\Gamma_{B_4} = \frac{\alpha}{\pi}d\Omega \frac{8}{M_1^2} \left[(\Lambda_9+\Lambda_{10}) |f_+(q^2)|^2 + \Lambda_9 \mathrm{Re} \,[f_+(q^2) f_-^*(q^2)] - \Lambda_{10} |f_-(q^2)|^2 \right], \label{eq:dgb4}
\end{equation}
and
\begin{equation}
d\Gamma_{B_5} = \frac{\alpha}{\pi}d\Omega \frac{8}{M_1^2} \left[\Lambda_{11} |f_+(q^2)|^2 + 2\Lambda_{11} \mathrm{Re} \,[f_+(q^2) f_-^*(q^2)] + \Lambda_{11} |f_-(q^2)|^2 \right]. \label{eq:dgb5}
\end{equation}
Here $\Lambda_i$ ($i=1,\ldots,11$) are functions of $E$ and $E_2$ and contain triple integrals over the relevant angular variables. They are listed in Appendix \ref{app:tripleint} for completeness.

Equations (\ref{eq:dgbir}) and (\ref{eq:dgbic})-(\ref{eq:dgb5}) constitute the first partial result for the bremsstrahlung radiative corrections. The infrared divergence and the finite terms that come along with it have been properly identified.
As a result, the bremsstrahlung differential decay rate can be obtained by performing numerically the triple integrals that make up the functions $\Lambda_i$ at various points $(E,E_2)$ of the Dalitz plot. This could require a considerable effort in an actual analysis.

We can proceed further, however, and also compute analytically the integrals displayed in Eqs.~(\ref{eq:dgbic})-(\ref{eq:dgb5}). The main aim of having a fully analytical expression available relies on the fact that it can be useful in the reduction of computing time in a Monte Carlo simulation of an experimental setup. For this task, then, a considerable amount of work can be saved if we realize that such integrals can be expressed in terms of the $\theta_i$ functions originally introduced in the analysis of bremsstrahlung radiative corrections in hyperon semileptonic decays \cite{tun89,tun91,rfm97,jjt04}. For conciseness, one has
\begin{equation}
\Lambda_1 =  \frac{p_2l}{2} \left(E Y_3 - 2EY_2 + 2\theta_0 \right), \label{eq:lambda1_th}
\end{equation}
\begin{equation}
\Lambda_2 = \frac{p_2lE}{2M_1} \left[-2\theta_0 + \frac{1}{E} Z_1 + (E+E_\nu^0)Y_2 -  E_\nu^0Y_3 \right], \label{eq:lambda2_th}
\end{equation}
\begin{equation}
\Lambda_3 = \frac{p_2lE}{8M_1^2} \left[ (2p_2ly_0-m^2)Y_3 - 2p_2l\beta Y_1 - 2Z_2 + 2 E(1-\beta^2) \theta_0 \right], \label{eq:lambda3_th}
\end{equation}
\begin{eqnarray}
\Lambda_4 & = & \frac{p_2\beta}{2} \bigg[ (E_\nu^0-2E)[\theta_7-2E(\theta_3-\theta_4)]
+ (3EE_\nu^0 - p_2ly_0)\theta_3 -3E^2(\theta_3-\theta_4-\beta\theta_5) \nonumber \\
&  & \mbox{} + 2E\eta_0 -E(1-\beta^2)\left[2E_\nu^0 \theta_2-\theta_6 + 2E(\theta_2-\theta_3)\right] - EE_\nu^0\theta_4 -E \theta_0 - \frac12 \theta_9 \bigg],
\end{eqnarray}
\begin{eqnarray}
\Lambda_5 & = & \frac{p_2\beta}{2M_1} \bigg[ \left[m^2(E+E_\nu^0)-E(l^2+EE_\nu^0)\right] \theta_3
-\frac{m^2}{2} \theta_7 + E(l^2+p_2ly_0-m^2)\theta_4 \nonumber \\
&  & \mbox{} + E(E-E_\nu^0)\eta_0 + E^2 \theta_0 + \beta p_2 m^2 \theta_{11} - El^2 \theta_{10} + \frac{El}{2} \theta_{14} + Em^2 Y_2 - E \zeta_{11} \bigg],
\end{eqnarray}
\begin{eqnarray}
\Lambda_6 & = & \frac{p_2\beta}{8M_1^2} \bigg[ m^2(p_2ly_0-l^2-EE_\nu^0) \theta_3 +  E(m^2E_\nu^0-2Ep_2ly_0) \theta_4 + EE_\nu^0l(2l\theta_{10}-\theta_{14}) \nonumber \\
&  & \mbox{} + \frac{E}{2}[2EE_\nu^0-2l^2+p_2l(1-y_0)]\eta_0 - Em^2\theta_0 + 2(E^2\zeta_{10} - m^2\zeta_{11}) \bigg],
\end{eqnarray}
\begin{eqnarray}
\Lambda_7 & = & \frac{p_2l}{2M_1} \bigg[ \left[E(1-\beta^2)(E+E_\nu^0)+EE_\nu^0+l^2\right] \theta_3  -(m^2+2EE_\nu^0)\theta_4 - E(\theta_0-\eta_0) \nonumber \\
&  & - \frac{E}{2} (2l\beta \theta_{10} - \beta \theta_{14}) - \frac{E}{2} (1-\beta^2) \theta_7 + \zeta_{11} - lE_\nu^0\theta_5 \bigg],
\end{eqnarray}
\begin{eqnarray}
\Lambda_8 & = & \frac{p_2l}{4M_1^2} \bigg[ \left( \frac{p_2}{2} (1-y_0)-l\right)l\eta_0 + (Ep_2ly_0+E_\nu^0m^2) \theta_4 - E\zeta_{10} + m^2\theta_0 \nonumber \\
&  & \mbox{} + 2l^2p_2(y_0\theta_5 - Y_1) - m^2 \left[ E_\nu^0+\beta(l+p_2y_0) \right] \theta_3 + \frac12 E_\nu^0l (2l\theta_{10}-\theta_{14}) \bigg],
\end{eqnarray}
\begin{equation}
\Lambda_9 = \frac{p_2l}{2M_1} \left( p_2ly_0 \theta_4 - E_\nu^0\eta_0 - \zeta_{10} \right),
\end{equation}
\begin{equation}
\Lambda_{10} = \frac{p_2lE}{4M_1^2} \left[ E_\nu^0\eta_0 - p_2ly_0\theta_4 + \frac{\beta}{2}E_\nu^0(2l\theta_{10} - \theta_{14}) + \zeta_{10} \right],
\end{equation}
and
\begin{equation}
\Lambda_{11} = \frac{Ep_2l}{8M_1^2} \bigg[ \beta E_\nu^0(2l\theta_{10}-\theta_{14}) + \frac12 \left[2E_\nu^0-2\beta l+\beta p_2(1-y_0)\right] \eta_0 + 2\beta p_2l(y_0\theta_5 - Y_1) \bigg], \label{eq:lambda11_th}
\end{equation}
where the secondary functions $\eta_0$, $\zeta_{lm}$, $Z_r$, and $Y_s$, given in Ref.~\cite{jjt04}, depend on the $\theta_i$ functions.

Finally, a further simplification of Eqs.~(\ref{eq:lambda1_th})-(\ref{eq:lambda11_th}) yields a fairly compact form of the bremsstrahlung differential decay rate. It can be organized as
\begin{equation}
d\Gamma_B = \frac{\alpha}{\pi} d\Omega \left( A_0 \theta_1 + A_B \right), \label{eq:dgbfinal}
\end{equation}
where
\begin{equation}
A_B = A_1^{(B)} |f_+(q^2)|^2 + A_2^{(B)} \mathrm{Re} \,[f_+(q^2) f_-^*(q^2)] + A_3^{(B)} |f_-(q^2)|^2, \label{eq:Ab}
\end{equation}
and the functions $A_i^{(B)}$ read
\begin{eqnarray}
\frac{M_1^2}{4p_2l}A_1^{(B)} & = & \left[\frac{2E_2}{M_1}+\frac{5E}{M_1}-\frac{E_2E}{M_1^2}-\frac{E^2}{M_1^2}(1+\beta^2) + \frac{p_2l}{2M_1^2}(1-y_0)\right] \eta_0 + \left[1-\frac{2E}{M_1} + \frac{3}{4}\frac{m^2}{M_1^2} \right] \theta_0 \nonumber \\
&  & \mbox{} - \frac{m^2}{E^2} \left[ 2E_\nu^0 + \frac{m^2}{M_1} \right]
\theta_2
+ \bigg[ E_\nu^0 - 3E - p_2\beta y_0 \nonumber \\
&  & \mbox{} + \frac{m^2}{E} \left[3-\frac{2E_2}{M_1}+\frac{m^2}{M_1^2}-\frac34 \frac{E(E+E_\nu^0)+p_2ly_0)}{M_1^2} \right] \bigg] \theta_3 \nonumber \\
&  & \mbox{} + \left[ E_\nu^0+\frac{2E_2E}{M_1}+\frac{4E^2}{M_1}+\frac{2p_2ly_0}{M_1}-\frac{m^2}{4M_1}\Bigg[9+\frac{3E_2}{M_1}+\frac{4E}{M_1}\Bigg]\right] \theta_4 \nonumber \\
&  & \mbox{} + \left[ 4+\frac{2E}{M_1} + \frac{2p_2ly_0}{M_1^2} - \frac{m^2}{4M_1^2} \right] l\theta_5 + \frac{m^2}{E^2} \theta_6  + \left[-2 + \frac{E_\nu^0}{E} - \frac{m^2}{M_1E} \right] \theta_7 - \frac{1}{2E} \theta_9 \nonumber \\
&  & \mbox{} + \frac{4l^2E_\nu^0}{M_1^2} \theta_{10}+ \frac{2p_2l}{M_1} \theta_{12} - \frac{2p_2l}{M_1} \theta_{13} + \frac{l(E+E_2)}{M_1^2} \theta_{14} - \frac{2p_2l^2}{M_1^2} \theta_{19} + \frac{2l^3}{M_1^2} \theta_{20},
\end{eqnarray}
\begin{eqnarray}
\frac{M_1^3}{m^2}\frac{1}{4p_2l} A_2^{(B)} & = & -\frac{1}{2M_1} \theta_0 + \frac{m^2}{E^2} \theta_2 - \left[ \frac32 + \frac{m^2}{EM_1} + \frac{E_2-p_2\beta y_0}{2M_1} \right] \theta_3 + \left[ \frac12 + \frac{2E+E_2}{2M_1} \right] \theta_4 \nonumber \\
&  & \mbox{} + \frac{l}{2M_1} \theta_5,
\end{eqnarray}
and
\begin{eqnarray}
A_3^{(B)} & = & \frac{p_2lm^2}{M_1^4} \left[ -\theta_0 + (M_1-E_2+\beta p_2y_0)\theta_3 - (M_1-E_2)\theta_4 - l\theta_5 \right].
\end{eqnarray}

Expression (\ref{eq:dgbfinal}) constitutes our second partial result for the bremsstrahlung differential decay rate of $K_{l3}^\pm$ decays, restricted to the TBR of the Dalitz plot. In spite of the lengthy expressions it consists of, its final form is easy to handle. Both infrared-divergent and infrared-convergent terms that appear in it have been rigorously identified and appropriately extracted. Let us now discuss the series of cross-checks performed on the functions $\Lambda_i$.

\subsection{Numerical cross-checks}

At this stage, we consider pertinent to cross-check the functions $\Lambda_i$ in their analytical forms (\ref{eq:lambda1_th})-(\ref{eq:lambda11_th}) against their counterparts (\ref{eq:lambda1_num})-(\ref{eq:lambda11_num}) with the triple integrals indicated, in order to ensure that our results are correct. These cross-checks consist in computing numerically the triple integrals and then contrasting the outputs with the direct evaluations of the corresponding analytical expressions at various points $(E,E_2)$ of the Dalitz plot. As an example, let us define the quantity
\begin{equation}
\mathcal{J} = \frac{p_2l}{2\pi} \beta^2\int_{-1}^1 dx \frac{1-x^2}{(1-\beta x)^2} \int_0^{2\pi}d\varphi_k\int_{-1}^{y_0} dy \frac{F}{4D^2} \frac{B_1}{\omega^2}, \label{eq:inum}
\end{equation}
where $B_1$ is given by Eq.~(\ref{eq:b1}). Its analytical counterpart is
\begin{equation}
\mathcal{J} = \frac{8}{M_1^2}(\Lambda_1+\Lambda_2+\Lambda_3), \label{eq:ith}
\end{equation}
with $\Lambda_r$ above given by Eqs.~(\ref{eq:lambda1_th})-(\ref{eq:lambda3_th}). We present in Table \ref{t:L1e} the evaluation of expressions (\ref{eq:inum}) and (\ref{eq:ith}) for $K_{e3}^\pm$ decays. We observe a very good agreement, entry by entry, between the two forms. Of course, we have also cross-checked the other $\Lambda_i$ functions and found very good agreements, too. We have repeated the exercise for $K_{\mu 3}^\pm$ decays and also found an excellent match, so there is no need to reproduce these outputs here.

\begingroup
\squeezetable
\begin{table}
\caption{\label{t:L1e}Values of $\mathcal{J}$ (see the text) in the TBR of the process $K^+ \to \pi^0 + e^+ + \nu_e$ by (a) performing numerically the triple integration in Eq.~(\ref{eq:inum}) and (b) evaluating  straightforwardly Eq.~(\ref{eq:ith}). The energies $E$ and $E_2$ are given in $\textrm{GeV}$ and $\mathcal{J}$ is dimensionless.}
\begin{ruledtabular}
\begin{tabular}{crrrrrrrrr}
$E_2\backslash E$ & $ 0.0123$ & $ 0.0370$ & $ 0.0617$ & $ 0.0864$ & $ 0.1111$ & $ 0.1358$ & $ 0.1604$ & $ 0.1851$ & $ 0.2098$ \\
\hline
 & & & & & (a) & & & \\
$ 0.2592$ & $ 0.3793$ & $-0.5527$ & $-2.0511$ & $-3.6155$ & $-4.9209$ & $-5.7042$ & $-5.7307$ & $-4.7722$ & $-2.5782$ \\
$ 0.2468$ &           & $ 0.7234$ & $-0.8941$ & $-2.6557$ & $-4.1663$ & $-5.1285$ & $-5.2902$ & $-4.4211$ & $-2.2974$ \\
$ 0.2345$ &           & $ 2.0511$ & $ 0.3903$ & $-1.5012$ & $-3.1741$ & $-4.3068$ & $-4.6362$ & $-3.9305$ & $-1.9761$ \\
$ 0.2222$ &           &           & $ 1.6926$ & $-0.3135$ & $-2.1392$ & $-3.4412$ & $-3.9464$ & $-3.4197$ & $-1.6509$ \\
$ 0.2098$ &           &           & $ 2.9937$ & $ 0.8784$ & $-1.0965$ & $-2.5672$ & $-3.2501$ & $-2.9063$ & $-1.3252$ \\
$ 0.1975$ &           &           &           & $ 2.0634$ & $-0.0593$ & $-1.6975$ & $-2.5578$ & $-2.3960$ & $-0.9997$ \\
$ 0.1851$ &           &           &           &           & $ 0.9642$ & $-0.8405$ & $-1.8758$ & $-1.8919$ & $-0.6739$ \\
$ 0.1728$ &           &           &           &           & $ 1.9645$ & $-0.0048$ & $-1.2106$ & $-1.3972$ & $-0.3461$ \\
$ 0.1604$ &           &           &           &           &           & $ 0.7955$ & $-0.5726$ & $-0.9158$ & $-0.0113$ \\
$ 0.1481$ &           &           &           &           &           &           & $ 0.0153$ & $-0.4537$ &           \\
\hline
 & & & & & (b) & & & \\
$ 0.2592$ & $ 0.3793$ & $-0.5527$ & $-2.0511$ & $-3.6155$ & $-4.9209$ & $-5.7042$ & $-5.7307$ & $-4.7720$ & $-2.5780$ \\
$ 0.2468$ &           & $ 0.7234$ & $-0.8941$ & $-2.6557$ & $-4.1663$ & $-5.1285$ & $-5.2901$ & $-4.4210$ & $-2.2971$ \\
$ 0.2345$ &           & $ 2.0511$ & $ 0.3903$ & $-1.5012$ & $-3.1741$ & $-4.3068$ & $-4.6362$ & $-3.9304$ & $-1.9760$ \\
$ 0.2222$ &           &           & $ 1.6926$ & $-0.3135$ & $-2.1392$ & $-3.4412$ & $-3.9463$ & $-3.4196$ & $-1.6507$ \\
$ 0.2098$ &           &           & $ 2.9937$ & $ 0.8784$ & $-1.0965$ & $-2.5672$ & $-3.2501$ & $-2.9062$ & $-1.3251$ \\
$ 0.1975$ &           &           &           & $ 2.0634$ & $-0.0593$ & $-1.6975$ & $-2.5578$ & $-2.3959$ & $-0.9996$ \\
$ 0.1851$ &           &           &           &           & $ 0.9642$ & $-0.8405$ & $-1.8757$ & $-1.8919$ & $-0.6739$ \\
$ 0.1728$ &           &           &           &           & $ 1.9645$ & $-0.0048$ & $-1.2106$ & $-1.3972$ & $-0.3461$ \\
$ 0.1604$ &           &           &           &           &           & $ 0.7954$ & $-0.5726$ & $-0.9158$ & $-0.0113$ \\
$ 0.1481$ &           &           &           &           &           &           & $ 0.0153$ & $-0.4536$ &           \\
\end{tabular}
\end{ruledtabular}
\end{table}
\endgroup

\section{\label{sec:close}Final results and comparison with other calculations}

The differential decay rate of $K_{l3}^\pm$ decays in the variables $E$ and $E_2$, restricted to the TBR of the Dalitz plot and including radiative corrections to the order of $(\alpha/\pi)(q/M_1)$, is given by
\begin{equation}
d\Gamma(K_{l3}^\pm) = d\Gamma_V + d\Gamma_B. \label{eq:dGfirst}
\end{equation}
$d\Gamma_V$ is given by Eq.~(\ref{eq:dGVfinal}). For $d\Gamma_B$ two forms are available. In the first one the triple integration over the real photon variables remains to be performed numerically. It is given by the sum of Eqs.~(\ref{eq:dgbir}) to (\ref{eq:dgb5}), which are written in terms of the functions $\Lambda_i$, Eqs.~(\ref{eq:lambda1_num})-(\ref{eq:lambda11_num}), listed in Appendix \ref{app:tripleint}. We should point out, however, that the infrared divergence cancels exactly in the sum in Eq.~(\ref{eq:dGfirst}). The second form for $d\Gamma_B$ is given by Eq.~(\ref{eq:dgbfinal}) and is expressed in terms of the fully analytical functions $\Lambda_i$, Eqs.~(\ref{eq:lambda1_th})-(\ref{eq:lambda11_th}). Let us remark, however, that obtaining this latter form was feasible by using results introduced in the analysis of radiative corrections in hyperon semileptonic decays \cite{tun89,tun91,rfm97,jjt04}.

Our analytical result can thus be cast into the compact form
\begin{equation}
d\Gamma(K_{l3}^\pm) = C_K^2 \frac{G_F^2|V_{us}|^2}{32\pi^3}M_1^3 dEdE_2 \left[ A_0^\prime + \frac{\alpha}{\pi} A^\prime \right], \label{eq:dgtot}
\end{equation}
where $A^\prime$ comprises all the various contributions arising from radiative corrections, namely,
\begin{eqnarray}
A^\prime & = & A_0^\prime (\Phi_1 +\theta_1) + A_V^\prime \Phi_2 + A_B^\prime \nonumber \\
& = & A_1 |f_+^\prime(q^2,p_+\cdot l)|^2 + A_2 \mathrm{Re} \,[f_+^\prime(q^2,p_+\cdot l) {f_-^\prime}^*(q^2,p_+\cdot l)] + A_3 |f_-^\prime(q^2,p_+\cdot l)|^2, \label{eq:fcorr}
\end{eqnarray}
where the functions $A_j$ are implicitly defined in Eq.~(\ref{eq:fcorr}). Here $A_0^\prime$, $A_V^\prime$, and $A_B^\prime$, given by Eqs.~(\ref{eq:a0}), (\ref{eq:a0p}), and (\ref{eq:Ab}), respectively, are functions of $E$ and $E_2$ and, to a very good approximation, depend quadratically on the modified form factors (\ref{eq:mff}), and the primes on them are an indicator of this fact. Besides, $\Phi_1$ and $\Phi_2$ are given by Eqs.~(\ref{eq:phiE}) and (\ref{eq:phiEp}). For $\theta_1$ we can use either Eq.~(96) of Ref.~\cite{tun89} or Eq.~(\ref{eq:i0}) of the present work since they are equivalent.

We are now in a position of producing some numerical evaluations which, at the same time, will allow us to compare our outputs with the ones obtained within other approaches. In the introductory remarks, we pointed out that some treatments about radiative corrections in $K_{l3}$ decays are available in the literature \cite{ginsKl3pm,ginsKe3pmDP,ginsKe30,ginsKm3,bech,maya,cir,cir2,bytev,andre}. In the present paper, we put emphasis on the radiative corrections to the Dalitz plot of $K_{l3}^\pm$ decays so a suitable comparison can readily be performed with the results presented in Refs.~\cite{cir,cir2} for this particular case.

In Ref.~\cite{cir}, the spin-averaged decay distribution $\rho(y,z)$ for $K_{l3}$ decays is analyzed, where, in our notation, the variables $y$ and $z$ correspond to $y=2E/M_1$ and $z=2E_2/M_1$, respectively. Also, the uncorrected distribution $\rho^{(0)}(y,z)$, Eq.~(3.5) of Ref.~\cite{cir}, is expressed in terms of the kinematical densities $A_1^{(0)}(y,z)$, $A_2^{(0)}(y,z)$, and $A_3^{(0)}(y,z)$, Eq.~(3.8) of this reference, which in turn are written in terms of the variables $r_\ell = m^2/M_1^2$ and $r_\pi = M_2^2/M_1^2$.

The uncorrected distribution $\rho^{(0)}(y,z)$ corresponds to our $d\Gamma_0(E,E_2)$ of Eq.~(\ref{eq:drate0}), and the amplitudes $A_i^{(0)}(y,z)$ are \textit{algebraically} equivalent to our $A_i^{(0)}(E,E_2)$ given in Eqs.~(\ref{eq:a10})-(\ref{eq:a30}). A numerical evaluation of $A_1^{(0)}$ for $K_{e3}^+$ decay is presented in Table 3 of Ref.~\cite{cir2}, which can be compared with Table \ref{t:eAi0}(a) of the present paper. The agreement, entry by entry, is evident.\footnote{We anticipated performing this comparison so we adopted the same notation for the amplitudes $A_i^{(0)}$ and evaluated $d\Gamma_0(E,E_2)$ at the same values of $E$ and $E_2$ as in Ref.~\cite{cir2}.}

On the other hand, the analysis of radiative corrections to $K_{l3}$ decays presented in Refs.~\cite{cir,cir2} was performed in the context of the chiral perturbation theory to the order of $\mathcal{O}(p^4,(m_d-m_u)p^2,e^2p^2)$, with the inclusion of the photon and light leptons as active degrees of freedom. In summary, the density distribution $\rho(y,z)$ with radiative corrections, which is equivalent to $d\Gamma$ of Eq.~(\ref{eq:dgtot}), is written as \cite{cir}
\begin{equation}
\rho(y,z) = \mathcal{N} S_{EW}(M_\rho,M_Z) \left[A_1 |f_+(t,v)|^2 + A_2 [f_+(t,v) f_-(q^2)] + A_3 |f_-(t,v)|^2 \right], \label{eq:rhorc}
\end{equation}
where the effective form factors $f_\pm$ depend on $t=(p_1-p_2)^2$ and $v=(p_1-l)^2$,
\begin{equation}
\mathcal{N} = C_K^2\frac{G_F^2|V_{us}|^2M_1^5}{128\pi^3},
\end{equation}
and $S_{EW}$ is the short distance enhancement factor.

The effects of virtual photons are contained in the long distance component $\Gamma_c(v,m^2,M^2;M_\gamma)$ of loop amplitudes (which produces infrared and Coulomb singularities); it depends on $v$ and the masses of the charged lepton $m$ and the charged meson $M$ and has a logarithmic dependence arising from the infrared regulator $M_\gamma$. On the other hand, the contributions of real soft photons are obtained by virtue of the theorem of Low. Therefore, the kinematical densities to the order of $\alpha$ read
\begin{equation}
A_i(y,z) = A_i^{(0)}(y,z) \left[1 + \Delta^{\mathrm{IR}}(y,z)\right] + \Delta^{\mathrm{IB}}(y,z). \label{eq:aicorr}
\end{equation}
Here $\Delta^\mathrm{IR}(y,z)$ contains $\Gamma_c(v,m^2,M^2;M_\gamma)$ and $|\mathcal{M}^\gamma_{(-1)}|^2$, where $\mathcal{M}^\gamma=\mathcal{M}^\gamma_{(-1)}+\mathcal{M}^\gamma_{(0)}$ is the radiative amplitude and $\mathcal{M}^\gamma_{(-1)}$ and $\mathcal{M}^\gamma_{(0)}$ contain terms of orders $1/k$ and $(k)^0$, respectively. Similarly, $\Delta^{\mathrm{IB}}(y,z)$ comprises the additional terms of $|\mathcal{M^\gamma}|^2$. From (\ref{eq:aicorr}), the radiative corrections to the Dalitz plot of $K_{l3}$ decays obtained in Refs.~\cite{cir,cir2} are
\begin{equation}
A_i(y,z)-A_i^{(0)}(y,z) = A_i^{(0)}(y,z) \Delta^{\mathrm{IR}}(y,z) + \Delta^{\mathrm{IB}}(y,z). \label{eq:e100}
\end{equation}
These latter contributions should be equivalent to our $(\alpha/\pi)A_i$ of Eq.~(\ref{eq:fcorr}), which read
\begin{equation}
\frac{\alpha}{\pi}A_i = \frac{\alpha}{\pi} \left[ A_i^{(0)}(\Phi_1 +\theta_1) + A_i^{(V)} \Phi_2 + A_i^{(B)} \right]. \label{eq:e101}
\end{equation}

Checking that Eqs.~(\ref{eq:e100}) and (\ref{eq:e101}) are algebraically equivalent is an involved task beyond the scope of the present paper. We limit ourselves to performing a numerical comparison of the available pieces of information instead.
For this purpose, we display a few samples of numerical values of the radiative corrections $(\alpha/\pi)A_i$ of Eq.~(\ref{eq:e101}) for both $K_{e3}^\pm$ and $K_{\mu 3}^\pm$ decays in Tables \ref{t:ecrA102030} and \ref{t:mcrA102030}, respectively. As anticipated, for $K_{e3}^\pm$ decays, $(\alpha/\pi)A_2$ and $(\alpha/\pi)A_3$ are negligible, but they are not so for $K_{\mu 3}^\pm$ decays. Additionally, the numerical results displayed in Table 4 of Ref.~\cite{cir2} for $A_1(y,z)-A_1^{(0)}(y,z)$ of $K_{e3}^\pm$ are reproduced in Table \ref{t:tcir} of the present work to carry on a comparison with Table \ref{t:ecrA102030}(a). An inspection of these tables shows an overall good agreement at the first significant digit over most of the entries and even the agreement at the second digit is more evident at the lowest entries of each column. There are small differences, however, and they may be explained as due to the different approximations used.

\begingroup
\squeezetable
\begin{table}
\caption{\label{t:ecrA102030}Radiative correction $(\alpha/\pi)A_i$, Eq.~(\ref{eq:fcorr}), in the TBR of the process $K^+ \to \pi^0 + e^+ + \nu_e$. The entries correspond to (a) $(\alpha/\pi)A_1\times 10$, (b) $(\alpha/\pi)A_2\times 10^6$, and (c) $(\alpha/\pi)A_3\times 10^7$. The energies $E$ and $E_2$ are given in $\textrm{GeV}$.}
\begin{ruledtabular}
\begin{tabular}{crrrrrrrrr}
$E_2\backslash E$ & $ 0.0123$ & $ 0.0370$ & $ 0.0617$ & $ 0.0864$ & $ 0.1111$ & $ 0.1358$ & $ 0.1604$ & $ 0.1851$ & $ 0.2098$ \\
\hline
 & & & & & (a) & & & \\
$0.2592$ & $ 0.1533$ & $ 0.1880$ & $ 0.1462$ & $ 0.0668$ & $-0.0286$ & $-0.1220$ & $-0.1949$ & $-0.2246$ & $-0.1726$ \\
$0.2468$ &           & $ 0.1810$ & $ 0.1580$ & $ 0.0902$ & $ 0.0011$ & $-0.0905$ & $-0.1654$ & $-0.1998$ & $-0.1537$ \\
$0.2345$ &           & $ 0.1578$ & $ 0.1522$ & $ 0.0962$ & $ 0.0150$ & $-0.0718$ & $-0.1445$ & $-0.1792$ & $-0.1354$ \\
$0.2222$ &           &           & $ 0.1429$ & $ 0.0989$ & $ 0.0261$ & $-0.0551$ & $-0.1249$ & $-0.1590$ & $-0.1168$ \\
$0.2098$ &           &           & $ 0.1321$ & $ 0.1004$ & $ 0.0363$ & $-0.0391$ & $-0.1056$ & $-0.1389$ & $-0.0977$ \\
$0.1975$ &           &           &           & $ 0.1014$ & $ 0.0461$ & $-0.0233$ & $-0.0863$ & $-0.1186$ & $-0.0779$ \\
$0.1851$ &           &           &           &           & $ 0.0558$ & $-0.0075$ & $-0.0670$ & $-0.0979$ & $-0.0568$ \\
$0.1728$ &           &           &           &           & $ 0.0654$ & $ 0.0083$ & $-0.0474$ & $-0.0769$ & $-0.0333$ \\
$0.1604$ &           &           &           &           &           & $ 0.0243$ & $-0.0275$ & $-0.0550$ & $-0.0019$ \\
$0.1481$ &           &           &           &           &           &           & $-0.0070$ & $-0.0316$ &           \\
\hline
 & & & & & (b) & & & \\
$0.2592$ & $ 0.0540$ & $ 0.0227$ & $-0.0011$ & $-0.0189$ & $-0.0315$ & $-0.0392$ & $-0.0417$ & $-0.0383$ & $-0.0269$ \\
$0.2468$ &           & $ 0.0270$ & $ 0.0028$ & $-0.0161$ & $-0.0301$ & $-0.0393$ & $-0.0436$ & $-0.0422$ & $-0.0339$ \\
$0.2345$ &           & $ 0.0281$ & $ 0.0034$ & $-0.0162$ & $-0.0311$ & $-0.0412$ & $-0.0466$ & $-0.0468$ & $-0.0414$ \\
$0.2222$ &           &           & $ 0.0034$ & $-0.0169$ & $-0.0324$ & $-0.0434$ & $-0.0499$ & $-0.0516$ & $-0.0496$ \\
$0.2098$ &           &           & $ 0.0034$ & $-0.0176$ & $-0.0339$ & $-0.0458$ & $-0.0533$ & $-0.0567$ & $-0.0589$ \\
$0.1975$ &           &           &           & $-0.0185$ & $-0.0356$ & $-0.0483$ & $-0.0570$ & $-0.0623$ & $-0.0700$ \\
$0.1851$ &           &           &           &           & $-0.0373$ & $-0.0510$ & $-0.0610$ & $-0.0685$ & $-0.0844$ \\
$0.1728$ &           &           &           &           & $-0.0393$ & $-0.0541$ & $-0.0655$ & $-0.0759$ & $-0.1075$ \\
$0.1604$ &           &           &           &           &           & $-0.0576$ & $-0.0710$ & $-0.0855$ & $-0.2223$ \\
$0.1481$ &           &           &           &           &           &           & $-0.0788$ & $-0.1014$ &           \\
\hline
 & & & & & (c) & & & \\
$0.2592$ & $ 0.0081$ & $ 0.0049$ & $ 0.0027$ & $ 0.0007$ & $-0.0011$ & $-0.0031$ & $-0.0054$ & $-0.0085$ & $-0.0143$ \\
$0.2468$ &           & $ 0.0189$ & $ 0.0122$ & $ 0.0064$ & $ 0.0007$ & $-0.0055$ & $-0.0129$ & $-0.0232$ & $-0.0424$ \\
$0.2345$ &           & $ 0.0338$ & $ 0.0226$ & $ 0.0128$ & $ 0.0031$ & $-0.0075$ & $-0.0202$ & $-0.0381$ & $-0.0725$ \\
$0.2222$ &           &           & $ 0.0333$ & $ 0.0194$ & $ 0.0057$ & $-0.0094$ & $-0.0278$ & $-0.0539$ & $-0.1056$ \\
$0.2098$ &           &           & $ 0.0440$ & $ 0.0260$ & $ 0.0081$ & $-0.0117$ & $-0.0361$ & $-0.0709$ & $-0.1432$ \\
$0.1975$ &           &           &           & $ 0.0325$ & $ 0.0102$ & $-0.0145$ & $-0.0452$ & $-0.0898$ & $-0.1883$ \\
$0.1851$ &           &           &           &           & $ 0.0118$ & $-0.0181$ & $-0.0556$ & $-0.1115$ & $-0.2473$ \\
$0.1728$ &           &           &           &           & $ 0.0126$ & $-0.0230$ & $-0.0682$ & $-0.1377$ & $-0.3407$ \\
$0.1604$ &           &           &           &           &           & $-0.0299$ & $-0.0844$ & $-0.1726$ & $-0.7783$ \\
$0.1481$ &           &           &           &           &           &           & $-0.1090$ & $-0.2302$ &           \\
\end{tabular}
\end{ruledtabular}
\end{table}
\endgroup

\begingroup
\squeezetable
\begin{table}
\caption{\label{t:mcrA102030}Radiative correction $(\alpha/\pi)A_i$, Eq.~(\ref{eq:fcorr}), in the TBR of the process $K^+ \to \pi^0 + \mu^+ + \nu_\mu$. The entries correspond to (a) $(\alpha/\pi)A_1\times 10^2$, (b) $(\alpha/\pi)A_2\times 10^3$, and (c) $(\alpha/\pi)A_3\times 10^3$. The energies $E$ and $E_2$ are given in $\textrm{GeV}$.}
\begin{ruledtabular}
\begin{tabular}{crrrrrrrrr}
$E_2\backslash E$ & $ 0.1131$ & $ 0.1280$ & $ 0.1429$ & $ 0.1578$ & $ 0.1727$ & $ 0.1876$ & $ 0.2025$ & $ 0.2174$ & $ 0.2322$ \\
\hline
 & & & & & (a) & & & \\
$0.2480$ &           &           &           &           & $-0.3524$ & $-0.2563$ & $-0.2278$ & $-0.1927$ & $-0.1099$ \\
$0.2361$ &           & $ 0.0248$ & $ 0.0020$ & $-0.0398$ & $-0.0880$ & $-0.1331$ & $-0.1635$ & $-0.1626$ & $-0.0952$ \\
$0.2242$ & $ 0.0933$ & $ 0.0777$ & $ 0.0417$ & $-0.0070$ & $-0.0607$ & $-0.1106$ & $-0.1452$ & $-0.1473$ & $-0.0809$ \\
$0.2123$ & $ 0.0887$ & $ 0.0782$ & $ 0.0466$ & $ 0.0014$ & $-0.0498$ & $-0.0980$ & $-0.1315$ & $-0.1332$ & $-0.0661$ \\
$0.2004$ & $ 0.0695$ & $ 0.0670$ & $ 0.0423$ & $ 0.0029$ & $-0.0435$ & $-0.0880$ & $-0.1190$ & $-0.1193$ & $-0.0510$ \\
$0.1885$ &           & $ 0.0512$ & $ 0.0344$ & $ 0.0019$ & $-0.0389$ & $-0.0789$ & $-0.1068$ & $-0.1053$ &           \\
$0.1766$ &           &           & $ 0.0248$ & $-0.0004$ & $-0.0350$ & $-0.0702$ & $-0.0946$ & $-0.0910$ &           \\
$0.1647$ &           &           & $ 0.0144$ & $-0.0032$ & $-0.0315$ & $-0.0616$ & $-0.0824$ & $-0.0766$ &           \\
$0.1528$ &           &           &           & $-0.0063$ & $-0.0280$ & $-0.0530$ & $-0.0701$ & $-0.0651$ &           \\
$0.1409$ &           &           &           &           &           & $-0.0448$ & $-0.0600$ &           &           \\
\hline
 & & & & & (b) & & & \\
$0.2480$ &           &           &           &           & $-0.2690$ & $-0.1865$ & $-0.1568$ & $-0.1271$ & $-0.0795$ \\
$0.2361$ &           & $ 0.0148$ & $-0.0315$ & $-0.0767$ & $-0.1136$ & $-0.1398$ & $-0.1527$ & $-0.1490$ & $-0.1263$ \\
$0.2242$ & $ 0.1431$ & $ 0.0611$ & $-0.0079$ & $-0.0662$ & $-0.1136$ & $-0.1490$ & $-0.1710$ & $-0.1779$ & $-0.1804$ \\
$0.2123$ & $ 0.1527$ & $ 0.0680$ & $-0.0054$ & $-0.0686$ & $-0.1213$ & $-0.1626$ & $-0.1915$ & $-0.2090$ & $-0.2497$ \\
$0.2004$ & $ 0.1556$ & $ 0.0696$ & $-0.0068$ & $-0.0739$ & $-0.1310$ & $-0.1775$ & $-0.2136$ & $-0.2434$ & $-0.3671$ \\
$0.1885$ &           & $ 0.0701$ & $-0.0094$ & $-0.0800$ & $-0.1415$ & $-0.1936$ & $-0.2377$ & $-0.2830$ &           \\
$0.1766$ &           &           & $-0.0124$ & $-0.0869$ & $-0.1531$ & $-0.2114$ & $-0.2651$ & $-0.3326$ &           \\
$0.1647$ &           &           & $-0.0159$ & $-0.0947$ & $-0.1662$ & $-0.2321$ & $-0.2986$ & $-0.4054$ &           \\
$0.1528$ &           &           &           & $-0.1044$ & $-0.1827$ & $-0.2590$ & $-0.3470$ & $-0.5828$ &           \\
$0.1409$ &           &           &           &           &           & $-0.3075$ & $-0.4629$ &           &           \\
\hline
 & & & & & (c) & & & \\
$0.2480$ &           &           &           &           & $-0.0058$ & $-0.0056$ & $-0.0064$ & $-0.0076$ & $-0.0100$ \\
$0.2361$ &           & $ 0.0006$ & $-0.0017$ & $-0.0041$ & $-0.0066$ & $-0.0093$ & $-0.0126$ & $-0.0172$ & $-0.0282$ \\
$0.2242$ & $ 0.0087$ & $ 0.0048$ & $ 0.0011$ & $-0.0027$ & $-0.0069$ & $-0.0116$ & $-0.0175$ & $-0.0264$ & $-0.0495$ \\
$0.2123$ & $ 0.0144$ & $ 0.0097$ & $ 0.0047$ & $-0.0006$ & $-0.0065$ & $-0.0134$ & $-0.0224$ & $-0.0364$ & $-0.0773$ \\
$0.2004$ & $ 0.0208$ & $ 0.0151$ & $ 0.0088$ & $ 0.0019$ & $-0.0059$ & $-0.0153$ & $-0.0277$ & $-0.0477$ & $-0.1254$ \\
$0.1885$ &           & $ 0.0210$ & $ 0.0132$ & $ 0.0046$ & $-0.0054$ & $-0.0174$ & $-0.0338$ & $-0.0611$ &           \\
$0.1766$ &           &           & $ 0.0176$ & $ 0.0071$ & $-0.0051$ & $-0.0202$ & $-0.0412$ & $-0.0787$ &           \\
$0.1647$ &           &           & $ 0.0220$ & $ 0.0094$ & $-0.0054$ & $-0.0241$ & $-0.0510$ & $-0.1053$ &           \\
$0.1528$ &           &           &           & $ 0.0109$ & $-0.0070$ & $-0.0305$ & $-0.0665$ & $-0.1706$ &           \\
$0.1409$ &           &           &           &           &           & $-0.0445$ & $-0.1054$ &           &           \\
\end{tabular}
\end{ruledtabular}
\end{table}
\endgroup

\begingroup
\squeezetable
\begin{table}
\caption{\label{t:tcir}Radiative correction $[A_1(y,z)-A_1^{(0)}(y,z)]\times 10$ presented in Table 4 of Ref.~\cite{cir2} and reproduced here for comparison with Table \ref{t:ecrA102030}(a) (see the text). The energies $E$ and $E_2$ are given in $\textrm{GeV}$.}
\begin{ruledtabular}
\begin{tabular}{crrrrrrrrr}
$E_2\backslash E$ & $ 0.0123$ & $ 0.0370$ & $ 0.0617$ & $ 0.0864$ & $ 0.1111$ & $ 0.1358$ & $ 0.1604$ & $ 0.1851$ & $ 0.2098$ \\
\hline
$0.2592$ & $ 0.1494$ & $ 0.1697$ & $ 0.1174$ & $ 0.0313$ & $-0.0670$ & $-0.1593$ & $-0.2275$ & $-0.2486$ & $-0.1841$ \\
$0.2468$ &           & $ 0.1708$ & $ 0.1364$ & $ 0.0610$ & $-0.0320$ & $-0.1236$ & $-0.1946$ & $-0.2213$ & $-0.1638$ \\
$0.2345$ &           & $ 0.1558$ & $ 0.1378$ & $ 0.0732$ & $-0.0128$ & $-0.1006$ & $-0.1704$ & $-0.1983$ & $-0.1440$ \\
$0.2222$ &           &           & $ 0.1356$ & $ 0.0821$ & $ 0.0036$ & $-0.0796$ & $-0.1474$ & $-0.1758$ & $-0.1240$ \\
$0.2098$ &           &           & $ 0.1321$ & $ 0.0898$ & $ 0.0190$ & $-0.0593$ & $-0.1248$ & $-0.1533$ & $-0.1035$ \\
$0.1975$ &           &           &           & $ 0.0971$ & $ 0.0341$ & $-0.0392$ & $-0.1021$ & $-0.1305$ & $-0.0822$ \\
$0.1851$ &           &           &           &           & $ 0.0490$ & $-0.0191$ & $-0.0794$ & $-0.1075$ & $-0.0597$ \\
$0.1728$ &           &           &           &           & $ 0.0639$ & $ 0.0010$ & $-0.0566$ & $-0.0841$ & $-0.0348$ \\
$0.1604$ &           &           &           &           &           & $ 0.0214$ & $-0.0333$ & $-0.0598$ & $-0.0020$ \\
$0.1481$ &           &           &           &           &           &           & $-0.0094$ & $-0.0340$ &           \\
\end{tabular}
\end{ruledtabular}
\end{table}
\endgroup

We can go further and provide a \textit{preliminary} expression for the decay rate of $K_{e3}^\pm$ decays by following the lines of Ref.~\cite{cir}, making use of the detailed determination of the relevant form factors performed in this reference.

From Eq.~(\ref{eq:dgtot}) and in analogy with Eq.~(7.5) of Ref.~\cite{cir} we have
\begin{equation}
\Gamma(K_{e3}^\pm) \sim C_K^2 \frac{G_F^2|V_{us}|^2}{128\pi^3}M_1^5 |f_+^{K^+\pi^0}(0)|^2 I(\tilde{\lambda}_+), \label{eq:gg}
\end{equation}
where
\begin{eqnarray}
I(\tilde{\lambda}_+) & = & \frac{4}{M_1^2} \int_m^{E_m}dE\int_{E_2^{\mathrm{min}}}^{E_2^{\mathrm{max}}}dE_2 \left[ A_1^{(0)} + \frac{\alpha}{\pi} A_1 \right]\left[1+\frac{q^2}{M_{\pi^\pm}^2} \tilde{\lambda}_+ \right]^2 \nonumber \\
& = & h_0 + h_1 \tilde{\lambda}_+ + h_2 \tilde{\lambda}_+^2, \label{eq:ilambda}
\end{eqnarray}
where the integration limits are given in (\ref{eq:EE2lim}) and the slope parameter $\tilde{\lambda}_+=0.0328\pm 0.0033$ has been estimated in Ref.~\cite{cir}. Notice that Eq.~(\ref{eq:gg}) does not contain the short distance enhancement factor $S_{EW}$ yet.

With no radiative corrections, the integral (\ref{eq:ilambda}) yields $h_0^{(0)}=0.0965$, $h_1^{(0)}=0.3337$, and $h_2^{(0)} =0.4618$, whereas the inclusion of radiative corrections yields $h_0=0.0958$, $h_1=0.3303$, and $h_2=0.4557$.
In order to compare under the same quotations with Ref.~\cite{cir}, we use $\tilde{\lambda}_+=0.030$, so radiative corrections cause a \textit{decrease} of $0.8$\% in the decay rate. This has to be compared with $h_0=0.09533$, $h_1=0.3287$, and $h_2=0.4535$ evaluated in this reference, which induces a decrease of 1.27\%.

\section{\label{sec:finalsec}Conclusions}

In this work we have obtained the radiative corrections to the Dalitz plot of $K_{l3}^\pm$ decays to the order of $(\alpha/\pi)(q/M_1)$, where $q$ is the momentum transfer and $M_1$ denotes the mass of the kaon. We have obtained a fully analytical expression which comprises contributions of both virtual and real photons, restricted to the three-body part of the allowed kinematical region. Despite its length, the analytical form obtained, Eq.~(\ref{eq:dgtot}), is quite simple and organized in a way that is easy to deal with. Among other properties, it contains all the terms of the order of $(\alpha/\pi)(q/M_1)$, it has no infrared divergences, it does not contain an ultraviolet cutoff, and it is not compromised by any model dependence of radiative corrections. As argued in Sec.~\ref{sec:vir}, the model dependence is absorbed into the already existing form factors by adding a function of $p_+\cdot l$ to $f_+(q^2)$ and another to $f_-(q^2)$. This, needless to say, is a theoretical problem and as such should be dealt with like that. The usefulness of Eq.~(\ref{eq:dgtot}), however, could be better appreciated when incorporated into a Monte Carlo simulation, since it may reduce the computational time required by the triple integrals.

We should emphasize that Eq.~(\ref{eq:dgtot}) is very useful for processes where the momentum transfer is not small and thus cannot be neglected. Thus, it is valid for any $M_{l3}^\pm$ decay, whether $M$ be $\pi^\pm$, $K^\pm$, $D^\pm$ or even $B^\pm$. To first order in $q$ it yields terms of the order of $(\alpha/\pi)(q/M_1)$ in the radiative corrections. The expected error by the omission of higher order terms is around $(\alpha/\pi)(q/M_1)^2 \approx 0.0012$ in $K^\pm$ and $D^\pm$ decays. Being conservative, if the accompanying factors amount to 1 order of magnitude increase, then we can estimate an upper bound to the theoretical uncertainty of 1.2\%. This should be acceptable with an experimental precision of 2\%-3\%. We envisage further improvements to our calculation by incorporating the effects of the FBR of the Dalitz plot and also the decay of a neutral kaon, which requires an extra effort \cite{fur}.

\begin{acknowledgments}
The authors are grateful to Consejo Nacional de Ciencia y Tecnolog{\'\i}a (Mexico) for partial support. J.J.T., A.M.,
and M.N.\ were partially supported by Comisi\'on de Operaci\'on y Fomento de Actividades Acad\'emicas (Instituto
Polit\'ecnico Nacional). They are also thankful for the warm hospitality extended to them at IF-UASLP, where part of this work was performed. R.F.-M.\ was also partially supported by Fondo de Apoyo a la Investigaci\'on (Universidad Aut\'onoma de San Luis Potos{\'\i}).
\end{acknowledgments}

\appendix

\section{\label{app:drbr}The bremsstrahlung differential decay rate and the infrared divergence}

In this appendix, we provide an outline of the method used to extract and isolate the infrared divergence in the present analysis. The method has been implemented in Ref.~\cite{tun89} to the analysis of baryon semileptonic decays so we have borrowed and adapted it to our analysis. Since most of the material presented here can be found in Secs.~III and IV of this reference, we have used the same conventions and notation.

The differential decay rate of process (\ref{eq:kl3gamma}) can be written as \cite{bd}
\begin{equation}
d\Gamma_B = \frac{1}{(2\pi)^8} \frac{1}{2M_1}\frac{mm_\nu}{4E_2EE_\nu\omega} d^3p_2\,d^3l\,d^3p_\nu\,d^3k \, \delta^4(p_1-p_2-l-p_\nu-k) \sum_{\textrm{spins},\epsilon}|\mathsf{M}_B|^2. \label{eq:diffdgbA}
\end{equation}

Proper integration over the variables involved in Eq.~(\ref{eq:diffdgbA}) requires that one chooses an orientation of the coordinate axes where the integrand acquires its simplest form. For this purpose, we orient the coordinate axes in such a way that $\mathbf{l}$, the three-momentum of the charged lepton, lies along the $+z$ direction, and $\mathbf{p}_2$, the three-momentum of the neutral pion, lies in the first or fourth quadrant of the $(x,z)$ plane.

Integrating (\ref{eq:diffdgbA}) over the neutrino three-momentum is an easy matter, so we are left with
\begin{equation}
d\Gamma_B = \frac{1}{(2\pi)^8} \frac{p_2l}{2M_1}\frac{mm_\nu}{4E_\nu}dEdE_2\, d(-\cos\theta_2)d\varphi_2\, d\Omega_l\,\frac{k^2}{\omega}dk\,d(-\cos\theta_k)d\varphi_k \, \delta(f(\cos\theta_2)) \sum_{\textrm{spins},\epsilon}|\mathsf{M}_B|^2, \label{eq:diffdgb1}
\end{equation}
where $\theta_2$ [$\theta_k$] and $\varphi_2$ [$\varphi_k$] are the polar and azimuthal angles of the pion [photon], respectively,
\begin{equation}
f(\cos\theta_2)=E_\nu^0 - \omega- \sqrt{(E_\nu^0-\omega)^2-C+a\cos\theta_2+b\sin\theta_2},
\end{equation}
and
\begin{equation}
C = (E_\nu^0-\omega)^2-p_2^2-l^2-k^2-2kl\cos\theta_k, \qquad a=2p_2(l+k\cos\theta_k), \qquad b=p_2k\sin\theta_k\cos\varphi_k,
\end{equation}
with $E_\nu^0=M_1-E-E_2$.

To proceed further we need to solve the equation $f(\cos\theta_2) = 0$ to find the zeros of the function $f(\cos\theta_2)$
in order to perform the integration over $\delta(f(\cos\theta_2))d(-\cos\theta_2)$ in Eq.~(\ref{eq:diffdgb1}). The use of energy and momentum conservation yields
\begin{subequations}
\label{eq:scth2}
\begin{equation}
\cos\theta_2^\pm = \frac{aC\pm b\sqrt{a^2+b^2-C^2}}{a^2+b^2},
\end{equation}
\begin{equation}
\sin\theta_2^\pm = \frac{bC\mp a\sqrt{a^2+b^2-C^2}}{a^2+b^2}.
\end{equation}
\end{subequations}
Since $0\leq \theta_2\leq \pi$, the signs of $\cos\theta_2^\pm$ and $\sin\theta_2^\pm$ are fixed unambiguously.

\begin{figure}[ht]
\scalebox{1.1}{\includegraphics{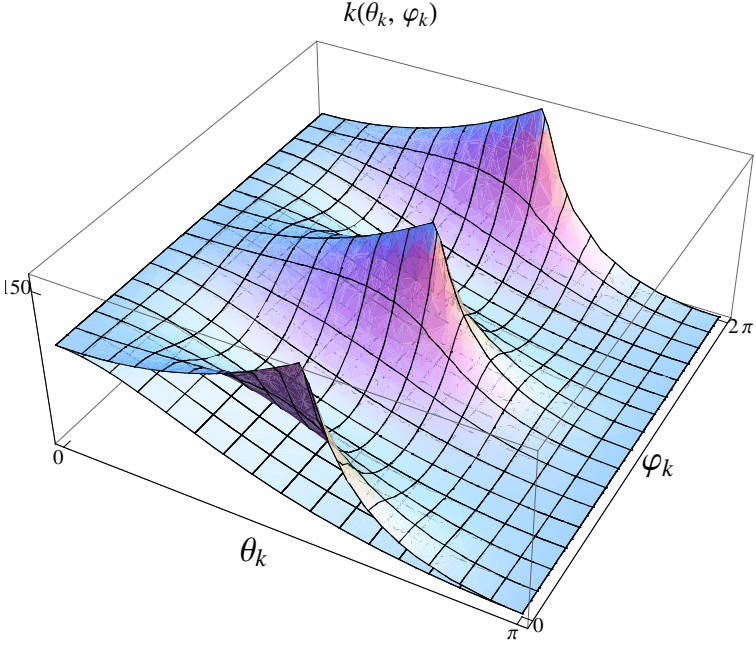}}
\caption{\label{fig:3d}Plot of $k_2$ and $k_4$ as functions of $\theta_k$ and $\varphi_k$ for $K_{e3}^+$ decay, at $E=111.1$ MeV and $E_2=222.2$ MeV. The uppermost surface represents $k_2$ so $k_4$ lies right below it.}
\end{figure}

\begin{figure}[ht]
\scalebox{1.1}{\includegraphics{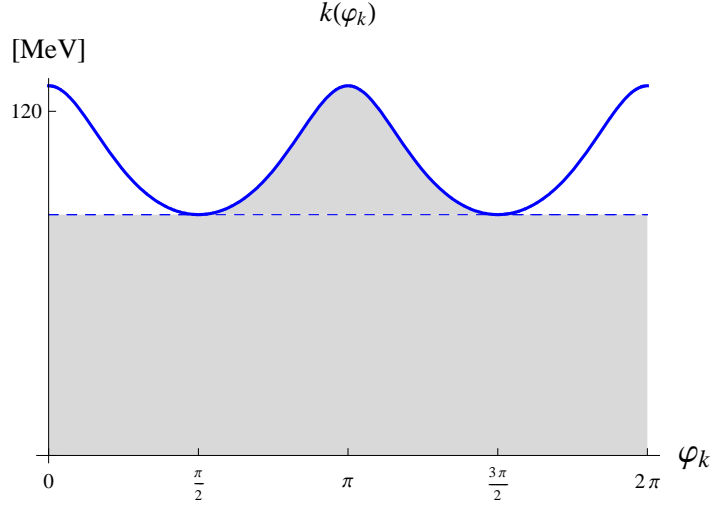}}
\caption{\label{fig:2d}Plot of $k_2$ and $k_4$ as functions of $\varphi_k$ for $\theta_k=\pi/3$, at the same values of $(E,E_2)$ as in Fig.~\ref{fig:3d}. The continuous and broken lines represent $k_2$ and $k_4$, respectively. The shaded area displays the accessible values of $k$ which lead to allowed values of $\cos\theta_2$.}
\end{figure}

Armed with Eqs.~(\ref{eq:scth2}) together with energy and momentum conservation relations, the physical values of the photon three-momentum can readily be obtained. First, let $k_{1,2}$ denote the zeros of the radicand in Eqs.~(\ref{eq:scth2}), and, next, let $k_{3,4}$ denote the values of $k$ that satisfy $C=\pm a$. The physical contents behind the plus and minus signs in the latter condition means that $\mathbf{l}$ and $\mathbf{p}_2$ are parallel and antiparallel, respectively, so the photon and neutrino three-momenta must rearrange accordingly to satisfy momentum conservation. We thus find
\begin{equation}
k_{1,2} = \frac{p_2c_1 -a_1b_1 \pm \sqrt{(p_2c_1-a_1b_1)^2+ (b_1^2-c_1^2)(d_1^2-a_1^2)}}{2(d_1^2-a_1^2)},
\end{equation}
and
\begin{equation}
k_{3,4} = \frac{b_1-2p_2(p_2\pm l)}{2(a_1\pm p_2\cos\theta_k)},
\end{equation}
where
\begin{equation}
a_1 = E_\nu^0+l\cos\theta_k, \qquad b_1 = {E_\nu^0}^2+p_2^2-l^2, \qquad c_1 = 2p_2E_\nu^0, \qquad d_1^2 = p_2^2(1-\sin^2\theta_k\sin^2\varphi_k).
\end{equation}
Throughout a careful and detailed analysis, one finds that $k_2$ and $k_4$ are the only physical values of $k$, whereas $k_1$ and $k_3$ are unphysical ones. In Fig.~\ref{fig:3d}, we have plotted $k_2$ and $k_4$ as functions of $\theta_k$ and $\varphi_k$ for $K_{e3}^+$ decay, at $E=111.1$ MeV and $E_2=222.2$ MeV for definiteness. In this figure, the uppermost surface corresponds to $k_2$ so $k_4$ lies just right below it.
Although $k_{2,4}$ are physical values, the condition $\sin\theta_2 \in [0,1]$ strongly constrains the accessible regions to them. In other words, in the region where $0 \leq k \leq k_4$ and $0\leq\varphi_k \leq 2\pi$, only $\sin\theta_2^-$ and $\cos\theta_2^-$ are allowed, whereas in the region where $k_4 \leq k \leq k_2$, both $\cos\theta_2^\pm$ and $\sin\theta_2^\pm$ are allowed, but this time $\varphi_k\in (\pi/2,3\pi/2)$. In Fig.~\ref{fig:2d}, we have plotted $k_2$ and $k_4$ as functions of $\varphi_k$ for $\theta_k=\pi/3$, at the same values of $(E,E_2)$ as above. The shaded area depicts
the accessible values of $k$ which lead to allowed values of $\cos\theta_2$.

All the above findings can be gathered together in order to express the bremsstrahlung differential decay rate as
\begin{eqnarray}
d\Gamma_B & = & \frac{1}{(2\pi)^6}\frac{p_2l}{2}\frac{mm_\nu}{M_1}dEdE_2 \int_{-1}^1 dx \bigg[ \nonumber \\
&  & \mbox{\hglue1.3truecm} \int_0^{2\pi}d\varphi_k\int_0^{k_4} dk \frac{k^2}{\omega} \left|\frac{\sin\theta_2^-}{a\sin\theta_2^--b\cos\theta_2^-} \right| \sum_{\textrm{spins},\epsilon}|\mathsf{M}_B|^2 \nonumber\\
&  & \mbox{\hglue1.3truecm} + \int_{\pi/2}^{3\pi/2}d\varphi_k\int_{k_4}^{k_2} dk\frac{k^2}{\omega} \left|\frac{\sin\theta_2^-}{a\sin\theta_2^--b\cos\theta_2^-} \right| \sum_{\textrm{spins},\epsilon}|\mathsf{M}_B|^2 \nonumber\\
&  & \mbox{\hglue1.3truecm} + \int_{\pi/2}^{3\pi/2}d\varphi_k\int_{k_4}^{k_2} dk\frac{k^2}{\omega} \left|\frac{\sin\theta_2^+}{a\sin\theta_2^+-b\cos\theta_2^+} \right| \sum_{\textrm{spins},\epsilon}|\mathsf{M}_B|^2 \bigg], \label{eq:dB1}
\end{eqnarray}
where $x=\cos\theta_k$ and $y=\cos\theta_2$. This form of $d\Gamma_B$, in terms of the set of variables $(k,\theta_2,\varphi_k)$ discussed in Sec.~\ref{sec:bremss}, is suitable for handling the infrared divergence although the involved factors in the integrand might complicate the evaluation of the integral to a great extent. There is another set, however, which allows one to handle the infrared-convergent contributions. It is expressed in the variables $(\theta_2,\theta_k,\varphi_k)$ through
\begin{equation}
k = \frac{F}{2D},
\end{equation}
where
\begin{eqnarray}
F & = & b_1 - 2p_2(p_2+l\cos\theta_k), \\
D & = & a_1 + p_2(\cos\theta_2\cos\theta_k + \sin\theta_2\sin\theta_k\cos\theta_k),
\end{eqnarray}
and after some algebraic manipulation we finally obtain
\begin{equation}
d\Gamma_B = \frac{1}{(2\pi)^6}\frac{p_2l}{2}\frac{mm_\nu}{M_1}dEdE_2 \int_{-1}^1 dx \int_0^{2\pi}d\varphi_k\int_{-1}^{y_0} dy \frac{F}{4D^2} \sum_{\textrm{spins},\epsilon}|\mathsf{M}_B|^2 . \label{eq:dB2}
\end{equation}
This latter form is the one actually used in the analysis of infrared-convergent contributions here.

\section{Integrals over the photon variables\label{app:tripleint}}

The various contributions $d\Gamma_{B_i}$ are constituted by the integrals listed below.
\begin{equation}
\Lambda_1 = \frac{p_2l}{4\pi} \beta^2 \int_{-1}^1 dx \frac{1-x^2}{(1-\beta x)^2} \int_{-1}^{y_0} dy \int_0^{2\pi} d\varphi_k \left[1 - \frac{E+l x}{D} \right], \label{eq:lambda1_num}
\end{equation}
\begin{equation}
\Lambda_2 = \frac{p_2l}{4\pi} \frac{\beta^2}{M_1} \int_{-1}^1 dx \frac{1-x^2}{(1-\beta x)^2} \int_{-1}^{y_0} dy \int_0^{2\pi} d\varphi_k \left[ \frac{E^2+E_\nu^0 l x+p_2 l y}{D} -E\right], \label{eq:lambda2_num}
\end{equation}
\begin{equation}
\Lambda_3 = \frac{p_2l}{4\pi} \frac{\beta^2}{4M_1^2} \int_{-1}^1 dx  \frac{1-x^2}{(1-\beta x)^2} \int_{-1}^{y_0} dy \int_0^{2\pi} d\varphi_k \left[ \frac{E(1-\beta x)(F-m^2)}{D} + m^2 \right], \label{eq:lambda3_num}
\end{equation}
\begin{eqnarray}
\Lambda_4 & = & \frac{p_2\beta}{4\pi} \int_{-1}^1 dx \frac{1}{1-\beta x} \int_{-1}^{y_0} dy \int_0^{2\pi} d\varphi_k  \frac{1}{D} \bigg\{ EE_\nu^0 - l^2 - p_2ly - \omega (E+l x) - E_\nu E (1-\beta x) \nonumber \\
&  & \mbox{} + (2 E_\nu - D) \left[\omega + E - \frac{E(1-\beta^2)}{1-\beta x} \right] \bigg\},
\end{eqnarray}
\begin{eqnarray}
\Lambda_5 & = & \frac{p_2\beta}{4\pi} \frac{1}{M_1} \int_{-1}^1 dx \frac{1}{1-\beta x} \int_{-1}^{y_0} dy \int_0^{2\pi} d\varphi_k \frac{1}{D} \bigg\{ -m^2 E_\nu^0 + \omega (p_2 l y+E^2+E_\nu^0 l x-E D)  \nonumber \\
&  & \mbox{} + E (1-\beta x) [\omega D+EE_\nu+l (p_2 y+l+\omega x)] - \left[1+\frac{E}{\omega}-\frac{E(1-\beta^2)}{\omega(1-\beta x)}\right] \nonumber \\
&  & \mbox{} \times [EE_\nu \omega (2-\beta x) + \omega l (l+p_2y+\omega x) - E \omega D] \bigg\},
\end{eqnarray}
\begin{eqnarray}
\Lambda_6 & = & \frac{p_2\beta}{4\pi} \frac{1}{4M_1^2} \int_{-1}^1 dx \frac{1}{1-\beta x} \int_{-1}^{y_0} dy \int_0^{2\pi} d\varphi_k \frac{1}{D} \bigg\{ m^2(EE_\nu^0+l^2+p_2ly_0) + EF\omega (1-\beta x) \nonumber \\
&  & \mbox{} - m^2E\omega (1-\beta x) + \frac{m^2}{2}F - E(1-\beta x)\bigg[(E+\omega)[F+2(EE_\nu+p_2ly+l^2+l\omega x)]
\nonumber \\
&  & \mbox{} - E_\nu [m^2+2E\omega (1-\beta x)]\bigg] + \bigg[1+\frac{E}{\omega}-\frac{E(1-\beta^2)}{\omega (1-\beta x)}\bigg] \nonumber \\
&  & \mbox{} \times \left[2\omega E (1-\beta x) (EE_\nu+l^2+p_2ly+l\omega x) - m^2 \omega D\right] \bigg\},
\end{eqnarray}
\begin{eqnarray}
\Lambda_7 & = & \frac{p_2l}{4\pi} \frac{1}{M_1} \int_{-1}^1 dx \int_{-1}^{y_0} dy \int_0^{2\pi} d\varphi_k \frac{1}{D} \bigg\{\omega E (1+\beta x)-\omega D-EE_\nu^0+l^2+p_2ly_0 \nonumber \\
&  & \mbox{} + \frac{1}{1-\beta x} \left[ EE_\nu(1-\beta^2) - \beta x \left[ (E+\omega)D-E_\nu^0lx - l^2- p_2ly_0\right]  \right] \bigg\},
\end{eqnarray}
\begin{eqnarray}
\Lambda_8 & = & \frac{p_2l}{4\pi} \frac{1}{M_1^2} \int_{-1}^1 dx \int_{-1}^{y_0} dy \int_0^{2\pi} d\varphi_k \frac{\omega}{2D} \bigg\{ -\frac{D}{\omega} \bigg[\left(1-\frac{1}{1-\beta x}\right)\left[m^2+2E\omega (1-\beta x) \right] \bigg] \nonumber \\
&  & \mbox{} - \frac{E(1-\beta^2)}{\omega (1-\beta x)} \left[EE_\nu^0+l^2+p_2ly_0-\omega E(1-\beta x)\right]
+ \frac{E_\nu^0}{\omega} m^2 + EE_\nu^0(1-\beta x) \nonumber \\
&  & \mbox{}  - E(M_1-E_2) - p_2ly_0 + (E+\omega)D \bigg\},
\end{eqnarray}
\begin{equation}
\Lambda_9 = \frac{p_2l}{4\pi} \frac{1}{M_1} \int_{-1}^1 dx \int_{-1}^{y_0} dy \int_0^{2\pi} d\varphi_k \omega \left[1 - \frac{2 E_\nu}{D} \right],
\end{equation}
\begin{eqnarray}
\Lambda_{10} = \frac{p_2l}{4\pi} \frac{1}{M_1^2} \int_{-1}^1 dx \int_{-1}^{y_0} dy \int_0^{2\pi} d\varphi_k \frac{\omega}{2D} \left[-ED + EE_\nu(2-\beta x) + p_2ly + l^2 + l\omega x \right],
\end{eqnarray}
\begin{eqnarray}
\Lambda_{11} = \frac{p_2l}{4\pi} \frac{E}{M_1^2} \int_{-1}^1 dx \int_{-1}^{y_0} dy \int_0^{2\pi} d\varphi_k
\frac{\omega}{2D} \left[ (1-\beta x)(E_\nu^0 - \omega - D) + D \right]. \label{eq:lambda11_num}
\end{eqnarray}


\begin{thebibliography}{99}

\bibitem{part}
  K.~Nakamura  [Particle Data Group],
  J.\ Phys.\ G {\bf 37}, 075021 (2010).

\bibitem{ag}
M.~Ademollo and R.~Gatto,
Phys.\ Rev.\ Lett.\  {\bf 13}, 264 (1964).

\bibitem{ginsKl3pm}
  E.~S.~Ginsberg,
  Phys.\ Rev.\  {\bf 142}, 1035 (1966).

\bibitem{ginsKe3pmDP}
  E.~S.~Ginsberg,
  Phys.\ Rev.\  {\bf 162}, 1570 (1967); {\bf 187}, 2280(E) (1969).

\bibitem{ginsKe30}
  E.~S.~Ginsberg,
  Phys.\ Rev.\  {\bf 171}, 1675 (1968); {\bf 174}, 2169(E) (1968); {\bf 187}, 2280(E) (1969).

\bibitem{ginsKm3}
  E.~S.~Ginsberg,
  Phys.\ Rev.\  D {\bf 1}, 229 (1970).

\bibitem{bech}
  T.~Becherrawy,
  Phys.\ Rev.\  D {\bf 1}, 1452 (1970).

\bibitem{maya}
A.~Garcia and M.~Maya,
Phys.\ Rev.\  D \textbf{23}, 2603 (1981).

\bibitem{cir}
  V.~Cirigliano, M.~Knecht, H.~Neufeld, H.~Rupertsberger and P.~Talavera,
  Eur.\ Phys.\ J.\  C {\bf 23}, 121 (2002).

\bibitem{cir2}
  V.~Cirigliano, H.~Neufeld and H.~Pichl,
  Eur.\ Phys.\ J.\  C {\bf 35}, 53 (2004).

\bibitem{bytev}
  V.~Bytev, E.~Kuraev, A.~Baratt and J.~Thompson,
  Eur.\ Phys.\ J.\  C {\bf 27}, 57 (2003)
  [Erratum-ibid.\  C {\bf 34}, 523 (2004)].

\bibitem{andre}
  T.~C.~Andre,
  Annals Phys.\  {\bf 322}, 2518 (2007).

\bibitem{sirlin}
A.~Sirlin,
Phys.\ Rev.\ \textbf{164}, 1767 (1967).

\bibitem{low}
F.~E.~Low,
Phys.\ Rev.\  \textbf{110}, 974 (1958).

\bibitem{chew}
H.\ Chew, Phys.\ Rev.\ \textbf{123}, 377 (1961).

\bibitem{juarez}
A.~Garcia and S.~R.~Juarez W.,
Phys.\ Rev.\ D \textbf{22}, 1132 (1980).

\bibitem{jl2009}
  C.~Juarez-Leon, A.~Martinez, M.~Neri, J.~J.~Torres, R.~Flores-Mendieta and A.~Garcia,
  Phys.\ Rev.\  D {\bf 79}, 057502 (2009), and references therein.

\bibitem{jj2006}
  J.~J.~Torres, M.~Neri, A.~Martinez, A.~Garcia and R.~Flores-Mendieta,
  Phys.\ Rev.\  D {\bf 74}, 077501 (2006), and references therein.

\bibitem{bd}
D.J.~Bjorken and S.D.~Drell, \textit{Relativistic Quantum Mechanics} (McGraw-Hill, New York, 1964).

\bibitem{bernard}
V.~Bernard, M.~Oertel, E.~Passemar and J.~Stern,
Phys.\ Rev.\  D {\bf 80}, 034034 (2009).

\bibitem{tun89}
D.~M.~Tun, S.~R.~Juarez W., and A.~Garcia,
Phys.\ Rev.\ D \textbf{40}, 2967 (1989).

\bibitem{rfm97}
R.~Flores-Mendieta, A.~Garcia, A.~Martinez, and J.~J.~Torres,
Phys.\ Rev.\ D \textbf{55}, 5702 (1997).

\bibitem{coester}
J.\ M.\ Jauch and F.\ Rohrlich, \textit{The Theory of Photons and Electrons} (Addison-Wesley, Reading MA, 1955). See Secs.\ 6-5 and 15-2.

\bibitem{tun91}
D.~M.~Tun, S.~R.~Juarez W., and A.~Garcia,
Phys.\ Rev.\ D \textbf{44}, 3589 (1991).

\bibitem{jjt04}
J.~J.~Torres, R.~Flores-Mendieta, M.~Neri, A.~Martinez and A.~Garcia,
Phys.\ Rev.\ D {\bf 70}, 093012 (2004).

\bibitem{fur}
C.\ Ju\'arez-Le\'on \textit{et.\ al.}\ work in progress.


\end{thebibliography}
\end{document}